\documentclass[letterpaper,twocolumn, 10pt]{article}
\usepackage{usenix,epsfig,endnotes}

\AtBeginDocument{%
  }
    
\usepackage{graphicx}
\usepackage{tipa}

\usepackage{tikz}
\usetikzlibrary{plotmarks}
\usetikzlibrary{positioning, shapes.geometric}
\usepackage{amsmath,soul}
\usepackage{xcolor}

\usepackage[english]{babel}
\usepackage{graphics}

\usepackage{amsmath}
\usepackage{url}
\usepackage{listings}
\usepackage{xcolor}
\usepackage{xspace}
\usepackage{booktabs}
\usepackage{tablefootnote}
\usepackage{algorithm}
\usepackage{algorithmic}
\usepackage{adjustbox}
\usepackage{comment}
\usepackage{numprint}
\usepackage{tcolorbox}
\usepackage{multirow}

\usepackage{colortbl}
\usepackage{xintexpr, xinttools}

\newcommand{\para}[1]{\smallskip \noindent \textbf{#1}}

\newcommand\eg{\emph{e.g.},\xspace}

\providecommand{\etal}{\emph{et al.}\xspace}

\expandafter\def\expandafter\UrlBreaks\expandafter{\UrlBreaks
  \do\a\do\b\do\c\do\d\do\e\do\f\do\g\do\h\do\i\do\j%
  \do\k\do\l\do\m\do\n\do\o\do\p\do\q\do\r\do\s\do\t%
  \do\u\do\v\do\w\do\x\do\y\do\z\do\A\do\B\do\C\do\D%
  \do\E\do\F\do\G\do\H\do\I\do\J\do\K\do\L\do\M\do\N%
  \do\O\do\P\do\Q\do\R\do\S\do\T\do\U\do\V\do\W\do\X%
  \do\Y\do\Z}

\providecommand{\UsePackageFor}[2]{ \ifx#2\undefined\usepackage{#1}\fi }

\UsePackageFor{amssymb}{\boxtimes}
\usepackage[normalem]{ulem}		
\usepackage{bbding}				
\usepackage{endnotes}			
\usepackage{hyperref}			
\usepackage{hyperendnotes}	
\usepackage{pdfcolfoot}			

\ifdefined\OrigFootnote\relax\else
	\newenvironment{FootnoteContent}{}{}
	\let\OrigFootnote\footnote
	\let\OrigFootnoteText\footnotetext
	\renewcommand{\footnotetext}[1]{\OrigFootnoteText{\begin{FootnoteContent}#1\end{FootnoteContent}}}
	\renewcommand{\footnote    }[1]{\OrigFootnote    {\begin{FootnoteContent}#1\end{FootnoteContent}}}
\fi

\definecolor{PurplePlum}{rgb}{0.1,0,0.55} 
\definecolor{Brown}{rgb}{0.5,.25,0}
\definecolor{Orange}{rgb}{1,.3,0}
\definecolor{Gray}{rgb}{.7,.7,.7}
\definecolor{DarkGreen}{rgb}{.1,.41,.1}

\newif\ifBleck

\newcommand\Colour[1] {\color{#1}}

\newcommand\PrintToCLinks{	
  {\Colour{blue}\mbox{
    \hyperlink{w1619}{\sf$\rightarrow$~top}\quad
    \hyperlink{w1031}{\sf$\rightarrow$~toc}\quad
    \hyperlink{w1148}{\sf$\rightarrow$~lof}\quad
    \hyperlink{GreenRoom}{\sf$\rightarrow$~gr}\quad
    \hyperlink{EndNotes}{\sf$\rightarrow$~en}\quad
    \hyperlink{Sargasso}{\sf$\rightarrow$~sg}\quad
    \hyperlink{Index}{\sf$\rightarrow$~idx}
  }}
}
\makeatletter 
\newcommand\ToCLinks{
  \ifx\@onlypreamble\@notprerr		
    \hypertarget{w1619}{}			
  \else
    \AtBeginDocument{\hypertarget{w1619}{}}	
  \fi

  \ifBleck\else	
    \ifdefined\cofoot
      \cofoot{\PrintToCLinks}
      \cefoot{\PrintToCLinks}
    \else
      \def\@oddfoot{\PrintToCLinks}
      \def\@evenfoot{\PrintToCLinks}
    \fi
 \fi
}
\makeatother

\newif\ifEndNotes 

\newcommand\FnSym{{\scriptsize\PencilLeftDown\kern.1em}}		
\newcommand\EnSym {{$\bigtriangledown$}}

\def\MarkupsHowto{} 
\newcommand{\MarkupsHowtoAdd}[1]{\expandafter\def\expandafter\MarkupsHowto\expandafter{\MarkupsHowto{}#1}} 

\newif\ifMarkupsHowtoPrinted 
\newif\ifSuppress 

\newcommand\MakeMarkups[3][.]{

     \Suppressfalse
     \ifBleck\Suppresstrue\fi
     \ifx0#1\Suppresstrue\fi
     \ifx1#1\Suppressfalse\fi
     
     \expandafter\providecommand\csname#2x\endcsname {} 
     \ifSuppress\expandafter\renewcommand\csname#2x\endcsname{\relax}\else
                       \expandafter\renewcommand\csname#2x\endcsname{#3}\fi
                       
     \expandafter\providecommand\csname#2\endcsname {} 
     \ifSuppress\expandafter\renewcommand\csname#2\endcsname[1]{##1}\else
                       \expandafter\renewcommand\csname#2\endcsname[1]{{\csname#2x\endcsname##1}}\fi

     \expandafter\providecommand\csname#2d\endcsname {} 
     \ifSuppress\expandafter\renewcommand\csname#2d\endcsname[1]{\relax}\else
                       \expandafter\renewcommand\csname#2d\endcsname[1]{{\csname#2x\endcsname\sout{##1}}}\fi
                       
     \expandafter\providecommand\csname#2r\endcsname {} 
     \ifSuppress\expandafter\renewcommand\csname#2r\endcsname[2]{{##2}}\else
                       \expandafter\renewcommand\csname#2r\endcsname[2]{\csname#2d\endcsname{##1} \csname#2\endcsname{##2}}\fi

     \expandafter\providecommand\csname#2i\endcsname {} 
     \ifSuppress\expandafter\renewcommand\csname#2i\endcsname[1]{\relax}\else
                       \expandafter\renewcommand\csname#2i\endcsname[1]{\csname#2\endcsname{##1}}\fi

     \expandafter\providecommand\csname#2t\endcsname {} 
     \ifSuppress\expandafter\renewcommand\csname#2t\endcsname[1]{\relax}\else
                       \expandafter\renewcommand\csname#2t\endcsname[1]{{\csname#2x\endcsname{\mbox{$\langle\!\langle$}##1{\csname#2x\endcsname\mbox{$\rangle\!\rangle$}}}}}\fi 

     \expandafter\providecommand\csname#2b\endcsname {} 
     \ifSuppress\expandafter\renewcommand\csname#2b\endcsname[1][empty]{\relax}\else 
                       \expandafter\renewcommand\csname#2b\endcsname[1][\empty]{\ifx\empty##1\empty
                       	\label{#2-bookmark} 
                              \marginpar [\raggedleft\csname#2\endcsname{{\footnotesize\fbox{#2 working here}}~$\Longrightarrow$}]
                                                {\csname#2\endcsname{$\Longleftarrow$~{\footnotesize\fbox{#2 working here}}}}
                       \else 
                       	\marginpar [\raggedleft\csname#2\endcsname{\ifx\empty##1\empty\else\fbox{\tiny\parbox{8em}{\raggedright##1}}~\fi$\Longrightarrow$}]
                                                {\csname#2\endcsname{$\Longleftarrow$\ifx\empty##1\empty\else~{\tiny\fbox{\parbox{8em}{\raggedright##1}}}\fi}}\fi}\fi

     \expandafter\providecommand\csname#2TD\endcsname {} 
     \ifSuppress\expandafter\renewcommand\csname#2TD\endcsname{\relax}\else
                       \expandafter\renewcommand\csname#2TD\endcsname{\csname#2\endcsname{\fbox{#2 to do}}}\fi

     \expandafter\providecommand\csname#2Bar\endcsname {} 
     \ifSuppress\expandafter\renewcommand\csname#2Bar\endcsname{\relax}\else
                       \expandafter\renewcommand\csname#2Bar\endcsname{\csname#2\endcsname{\scriptsize\XSolidBrush}}\fi

     \expandafter\providecommand\csname#2f\endcsname {} 
     \ifSuppress\expandafter\renewcommand\csname#2f\endcsname[2][]{\relax}\else
      \expandafter\renewcommand\csname#2f\endcsname[2][\empty]{ 
        {\mbox{\csname#2x\endcsname\tiny$\boxtimes$}\marginpar{\hsize1cm\csname#2x\endcsname\fbox{\FnSym\footnotemark}}\relax 
        \footnotetext{\csname#2x\endcsname##2}}}\fi

     \expandafter\providecommand\csname#2e\endcsname {}
     \ifSuppress\expandafter\renewcommand\csname#2e\endcsname[1]{\relax}\else%
      \expandafter\renewcommand\csname#2e\endcsname[1]{%
       \global\EndNotestrue
       \mbox{\scriptsize\csname#2x\endcsname$\boxtimes$}\relax%
       \marginpar{\hsize1cm\csname#2x\endcsname\fbox{\EnSym\endnotemark%
                          \hypertarget{ENmark\thepage-\theendnote}{}~\hyperlink{ENtext\thepage-\theendnote}{{\Colour{blue}$\downarrow$}}}%
       }%
       {
        \def\zz{\noexpand#3}%
        \edef\z{~{[Endnote \theendnote\ %
        on p.\noexpand\hypertarget{ENtext\thepage-\theendnote}{}\thepage%
                    ~\noexpand\hyperlink{ENmark\thepage-\theendnote}{{\noexpand\Colour{blue}$\uparrow$}}]}%
        }%
        \expandafter\endnotetext\expandafter{\z\vspace{2ex}\\ ##1\newpage}%
       }
      }\fi

     \expandafter\providecommand\csname#2n\endcsname {}
     \ifSuppress\expandafter\renewcommand\csname#2n\endcsname[1]{\relax}\else%
      \expandafter\renewcommand\csname#2n\endcsname[1]{%
       \global\EndNotestrue
    \marginpar{{\tiny\endnotemark}\hypertarget{ENmark\thepage-\theendnote}{}~\hyperlink{ENtext\thepage-\theendnote}{}}
       {
        \def\zz{\noexpand#3}%
        \edef\z{~{\zz[Endnote (deferred) 
        from p.\noexpand\hypertarget{ENtext\thepage-\theendnote}{}\thepage%
        ]}%
        }%
        \expandafter\endnotetext\expandafter{\z\vspace{2ex}\\ ##1\newpage}%
       }
      }\fi

     \expandafter\providecommand\csname#2fe\endcsname {} 
     \ifSuppress\expandafter\renewcommand\csname#2fe\endcsname[2][]{\relax}\else 
      \expandafter\renewcommand\csname#2fe\endcsname[2][]{ 
       \def\File{##1}\relax
       \ifx\File\empty\csname#2f\endcsname{##2}\else 
        \global\EndNotestrue 
        \mbox{\scriptsize\csname#2x\endcsname$\boxtimes$}
        \marginpar{\csname#2x\endcsname\fbox{\FnSym\footnotemark}}\relax
        \footnotetext{~\csname#2x\endcsname##2\
                             --- See [\EnSym\endnotemark\hypertarget{ENmark\thepage-\theendnote}{}
                             \kern-.2em\hyperlink{ENtext\thepage-\theendnote}{{\Colour{blue}$\downarrow$}}].}\relax
       { 
         \def\zz{\noexpand#3}
         \edef\z{~{\zz[Endnote~\thefootnote~on~p.\noexpand\hypertarget{ENtext\thepage-\theendnote}{}\thepage
                     ~\noexpand\hyperlink{ENmark\thepage-\theendnote}
                     {{\noexpand\Colour{blue}\kern-0.1em$\uparrow$}]}}
                     {\noexpand\footnotesize\noexpand\newline\noexpand\hspace*{2em} (~from file {\noexpand\tt\File.tex}~)}
         }    
         \expandafter\endnotetext\expandafter{\z~\par\input{##1}\newpage}
        } 
       \fi 
      } 
     \fi 

     \ifSuppress\relax\else\ifBleck\relax\else
      \MarkupsHowtoAdd{\par\csname#2t\endcsname{
       $\backslash$\texttt{#2}$\cdots$\ markups are in \textbf{this} colour\ifx#1..\else\ifx1#1.\else, e.g.\ for #1.\fi\fi
       \ifMarkupsHowtoPrinted\relax\else 
        \global\MarkupsHowtoPrintedtrue 
        \begin{quote}\begin{tabular}{l@{\hspace{2em}}p{.7\linewidth}}
         \multicolumn{2}{l}{\texttt{$\backslash$MakeMarkups\ifx#1.\relax\else[#1]\fi\{#2\}\{{\it$\langle$colour command\/$\rangle$}\}}
         				 --- Defines the macros below:}\\
             & see comments at \texttt{$\backslash$MakeMarkups} definition. \\[1ex]
         \texttt{$\backslash$#2\{$\langle$text$\rangle$\}} & Sets \texttt{$\langle$text$\rangle$} in \texttt{#2}'s colour. \\
         \texttt{$\backslash$#2x} & Changes to \texttt{#2}'s colour (until end of context). \\
         \texttt{$\backslash$#2d\{$\langle$text$\rangle$\}} & Sets \texttt{$\langle$text$\rangle$} in \texttt{#2}'s colour with a strikethrough (i.e.\ delete). \\
         \texttt{$\backslash$#2r\{$\langle$this$\rangle$\}\{$\langle$that$\rangle$\}} &
          Strikes through \texttt{$\langle$this$\rangle$} and inserts \texttt{$\langle$that$\rangle$} (i.e.\ replace). \\
         \texttt{$\backslash$#2f\{$\langle$text$\rangle$\}} & Meta-comment: puts \texttt{$\langle$text$\rangle$} in a \texttt{#2}-footnote with a {\tiny$\boxtimes$} in the main text. \\
         \texttt{$\backslash$#2t\{$\langle$text$\rangle$\}} & Use for meta when  \texttt{$\backslash$#2f} isn't allowed (``Not in outer-par mode.'') \\
         \texttt{$\backslash$#2b[$\langle$optional$\rangle$]} & Marginal pointer, with label for hyper-linking directly there. \\
         \texttt{$\backslash$#2e\{$\langle$text$\rangle$\}} & Puts \texttt{$\langle$text$\rangle$} in a \texttt{#2}-endnote with a (big) $\boxtimes$ in the main text. \\[.5ex]
         \texttt{$\backslash$#2n\{$\langle$text$\rangle$\}} & Like \texttt{$\backslash$#2e}
         except there's no reference from the main text. Good for ``decluttering''
         when you still want to have the footnote- or endnote texts as reminders. \\[.5ex]
         \texttt{$\backslash$#2fe[$\langle$this$\rangle$]\{$\langle$that$\rangle$\}} & Makes a \texttt{$\backslash$#2f\{$\langle$that$\rangle$\}} that refers to a \\
           & \texttt{$\backslash$#2e\{$\langle$contents of file this.tex$\rangle$\}}. \\ 
           & Without the optional argument, acts as \texttt{$\backslash$#2f\{$\langle$that$\rangle$\}}. \\[.5ex]
         \texttt{$\backslash$#2Bar} & Inserts ``burn after reading'' symbol \csname#2Bar\endcsname, meaning
          \begin{quote}\begin{itemize}\setlength\itemsep{0pt}
           \item If yours is the only \csname#2Bar\endcsname\ in this (presumably someone else's) footnote, and you are happy that the footnote has been addressed,
           go ahead and comment-out the whole footnote. (The \csname#2Bar\endcsname\ is their request for you to ``approve and remove''.)
           \item If you are not happy, delete only your \csname#2Bar\endcsname\ and follow-on in the footnote
            (in your colour, i.e.\ with \texttt{$\backslash$#2x}) saying why you are not happy.
           \item If you are happy, but there are others' burn-after-reading symbols as well as yours, just delete yours; the other people have not yet responded.
          \end{itemize}
          \end{quote}
          The idea is that when everyone's happy, the last person will comment-out the meta-text. \\[0.5ex]
         \texttt{$\backslash$#2TD} & Inserts {\csname#2TD\endcsname}\ . \\
        \end{tabular}\end{quote}
       \fi
      }}
     \fi\fi
}

\newif\ifNoGreenRoom
\newcommand\MakeGreenRoom {\ifBleck\relax\else\ifNoGreenRoom\relax\else
\newcommand\NewGRLabel[1] {\OldGRLabel{GreenRoom-##1}} 
 \newcommand\NewGRRef[1] 
 {\expandafter\ifx\csname r@GreenRoom-##1\endcsname\relax\OldGRRef{##1}\else\OldGRRef{GreenRoom-##1}\fi}
 \let\OldGRLabel\label \let\label\NewGRLabel
 \let\OldGRRef\ref \let\ref\NewGRRef
 \hrule
 ~\\\begin{center}\Huge \hypertarget{GreenRoom}{Green Room}
 \end{center}~\\
 \hrule
\fi\fi}

\newcommand\EndGreenRoom  {\ifBleck\relax\else\ifNoGreenRoom\relax\else
\let\label\OldGRLabel
\let\ref\OldGRRef
\fi\fi}

\newif\ifNoEndNotes

\newif\ifNoSargasso
\newcommand\MakeSargasso {
 \hypertarget{Sargasso}{}
 \newcommand\NewLabel[1] {\OldLabel{Sargasso-##1}} 
 \newcommand\NewRef[1] 
 {\expandafter\ifx\csname r@Sargasso-##1\endcsname\relax\OldRef{##1}\else\OldRef{Sargasso-##1}\fi}
 \let\OldLabel\label \let\label\NewLabel
 \let\OldRef\ref \let\ref\NewRef
\ifBleck\end{document}\else\ifNoSargasso
\relax
\else
  \hrule
  ~\\\begin{center}\Huge Sargasso
  \end{center}~\\
  \hrule
 \fi\fi
}

\newcommand\EndSargasso  {\ifBleck\relax\else\ifNoSargasso\relax\else
\let\label\OldLabel
\let\ref\OldRef
\fi\fi}

\newcommand\EndDocument {\ifBleck\end{document}\fi} 

\newcommand\Cite[2][\empty] {{\Colour{red}\ifx#1\empty[#2]\else[#2,~#1]\fi}}

\microtypecontext{spacing=nonfrench}
\begin{document}
\date{}

\newcommand\name{\textsc{Purl}\xspace}
\newcommand\cookiegraph{\textsc{CookieGraph}\xspace}
\newcommand\webgraph{WebGraph\xspace}

\newcommand*\fullcirc[1][1ex]{\tikz\fill (0,0) circle (#1);} 
\newcommand{\halfcirc}{
\tikz
    {
    \node (s1) [circle, fill=white, minimum size=1ex] {};
    \node      [semicircle, fill=black, 
                inner sep=0pt, outer sep=0pt, 
                minimum size=1ex,
                at={(s1.center)}, 
                rotate=90] {};
     }
}

\newcommand{\ats}{ATS\xspace}
\newcommand{\atses}{ATSes\xspace}
\newcommand{\nonats}{Non-ATS\xspace}
\newcommand{\nonatses}{Non-ATSes\xspace}
\newcommand{\unknown}{Unknown\xspace}

\newcommand{\accuracy}{98.74}
\newcommand{\precision}{98.62}
\newcommand{\recall}{98.87}

\newcommand{\atscookie}{first-party \ats cookie\xspace}
\newcommand{\nonatscookie}{first-party \nonats cookie\xspace}
\newcommand{\atscookies}{first-party \ats cookies\xspace}
\newcommand{\nonatscookies}{first-party \nonats cookies\xspace}

\newcommand{\cblockcrawl}{\texttt{3P-Blocked}\xspace}
\newcommand{\callowcrawl}{\texttt{3P-Allowed}\xspace}

\definecolor{lightgray}{rgb}{0.95, 0.95, 0.95}
\definecolor{darkgray}{rgb}{0.4, 0.4, 0.4}
\definecolor{editorGray}{rgb}{0.95, 0.95, 0.95}
\definecolor{editorOcher}{rgb}{1, 0.5, 0} 
\definecolor{editorGreen}{rgb}{0, 0.5, 0} 
\definecolor{orange}{rgb}{1,0.45,0.13}		
\definecolor{olive}{rgb}{0.17,0.59,0.20}
\definecolor{brown}{rgb}{0.69,0.31,0.31}
\definecolor{purple}{rgb}{0.38,0.18,0.81}
\definecolor{lightblue}{rgb}{0.1,0.57,0.7}
\definecolor{lightred}{rgb}{1,0.4,0.5}

\def\bluecolor{\color{blue}}

\lstdefinelanguage{JavaScript}{
  morekeywords={typeof, new, true, false, catch, function, return, null, catch, switch, var, if, in, while, do, else, case, break},
  morecomment=[s]{/*}{*/},
  morecomment=[l]//,
  morestring=[b]",
  morestring=[b]'
}

\lstdefinelanguage{HTML5}{
  language=html,
  sensitive=true,	
  alsoletter={<>=-},	
  morecomment=[s]{<!-}{-->},
  tag=[s],
  otherkeywords={
  >,
	<!DOCTYPE,
  </html, <html, <head, <title, </title, <style, </style, <link, </head, <meta, />,
	</body, <body,
	</div, <div, </div>, 
	</p, <p, </p>,
	</script, <script,
  <canvas, /canvas>, <svg, <rect, <animateTransform, </rect>, </svg>, <video, <source, <iframe, </iframe>, </video>, <image, </image>, <header, </header, <article, </article
  },
  ndkeywords={
  =,
  charset=, src=, id=, width=, height=, style=, type=, rel=, href=,
  fill=, attributeName=, begin=, dur=, from=, to=, poster=, controls=, x=, y=, repeatCount=, xlink:href=,
  margin:, padding:, background-image:, border:, top:, left:, position:, width:, height:, margin-top:, margin-bottom:, font-size:, line-height:,
  transform:, -moz-transform:, -webkit-transform:,
  animation:, -webkit-animation:,
  transition:,  transition-duration:, transition-property:, transition-timing-function:,
  }
}

\lstdefinestyle{htmlcssjs} {%
  basicstyle={\footnotesize\ttfamily},   
  frame=b,
  identifierstyle=\color{black},
  keywordstyle=\color{blue}\bfseries,
  ndkeywordstyle=\color{editorGreen}\bfseries,
  stringstyle=\color{editorOcher}\ttfamily,
  commentstyle=\color{brown}\ttfamily,
  language=HTML5,
  alsolanguage=JavaScript,
  alsodigit={.:;},	
  tabsize=2,
  showtabs=false,
  showspaces=false,
  showstringspaces=false,
  extendedchars=true,
  breaklines=true,
  literate=%
  {Ö}{{\"O}}1
  {Ä}{{\"A}}1
  {Ü}{{\"U}}1
  {ß}{{\ss}}1
  {ü}{{\"u}}1
  {ä}{{\"a}}1
  {ö}{{\"o}}1
}

\MakeMarkups[Shaoor]{S}{\Colour{DarkGreen}}
\MakeMarkups[Sandra]{A}{\Colour{purple}}
\MakeMarkups[Zubair]{Z}{\Colour{magenta}}
\MakeMarkups[Steven]{T}{\Colour{cyan}}
\MakeMarkups[Shaoor]{H}{\Colour{blue}}

\newcommand{\redline}{\raisebox{2pt}{\tikz{\draw[-,red,solid,line width = 3pt](0,0) -- (5mm,0);}}}
\newcommand{\orangeline}{\raisebox{2pt}{\tikz{\draw[-,orange,solid,line width = 3pt](0,0) -- (5mm,0);}}}
\newcommand{\greenline}{\raisebox{2pt}{\tikz{\draw[-,green,solid,line width = 3pt](0,0) -- (5mm,0);}}}

\newcommand{\blackline}{\raisebox{2pt}{\tikz{\draw[-,black,solid,line width = 3pt](0,0) -- (5mm,0);}}}
\newcommand{\grayline}{\raisebox{2pt}{\tikz{\draw[-,gray,solid,line width = 3pt](0,0) -- (5mm,0);}}}
\newcommand{\lightgrayline}{\raisebox{2pt}{\tikz{\draw[-,lightgray,solid,line width = 3pt](0,0) -- (5mm,0);}}}

\makeatletter 
\newcommand{\linebreakand}{%
  \end{@IEEEauthorhalign}
  \hfill\mbox{}\par
  \mbox{}\hfill\begin{@IEEEauthorhalign}
}
\makeatother 

\title {\name: Safe and Effective Sanitization of Link Decoration}

\author{
{\rm Shaoor Munir}\\
UC Davis \\
smunir@ucdavis.edu
\and
{\rm Patrick Lee}\\
UC Davis \\
pelee@ucdavis.edu
\and
{\rm Umar Iqbal}\\
Washington University in St. Louis \\
umar.iqbal@wustl.edu
\and
{\rm Zubair Shafiq}\\
UC Davis \\
zubair@ucdavis.edu
\and
{\rm Sandra Siby}\\
Imperial College London \\
s.siby@imperial.ac.uk
}

\maketitle

\begin{abstract}
	While privacy-focused browsers have taken steps to block third-party cookies and mitigate browser fingerprinting, novel tracking techniques that can bypass existing countermeasures continue to emerge.
Since trackers need to share information from the client-side to the server-side through link decoration regardless of the tracking technique they employ, a promising orthogonal approach is to detect and sanitize tracking information in decorated links.
To this end, we present \name (pronounced purel-l), a machine-learning approach that leverages a cross-layer graph representation of webpage execution to safely and effectively sanitize link decoration. 
Our evaluation shows that \name significantly outperforms existing countermeasures in terms of accuracy and reducing website breakage while being robust to common evasion techniques. 
\name's deployment on a sample of top-million websites shows that link decoration is abused for tracking on nearly three-quarters of the websites, often to share cookies, email addresses, and fingerprinting information.
\vspace{-10pt}
\end{abstract}

\section{Introduction}
\label{sec:introduction}
Web browsers and browser extensions are actively cracking down on new and emerging online tracking techniques. 
For example, almost all mainstream web browsers now already block or will soon block third-party cookies~\cite{safari-default-cookie, firefox-cookie-policy, GooglePrivacySandbox} and some privacy-focused browsers even deploy countermeasures against emerging tracking techniques, such as browser fingerprinting \cite{webkit-fingerprinting, firefox-fingerprinting, brave-privacy-updates}.
In response, online trackers continue to evolve and devise innovative tracking techniques that can bypass existing privacy-enhancing countermeasures \cite{Sanchez2021JourneyToCookies, ChenACM2021CookieSwapParty, fouad2022my, Munir23CookieGraphArxiv}. 
The key limitation of existing privacy-enhancing tools is that they aim to mitigate specific types of tracking (e.g., third-party cookies, first-party cookies, email addresses, canvas fingerprinting, AudioContext fingerprinting, CNAME cloaking) \cite{Englehardt16OpenWPMCCS,Dao2021CNAMECloaking,ChenACM2021CookieSwapParty,Englehardt18EmailPETS}. 
This approach works reasonably well to protect against known forms of tracking but fails against new or unknown forms of tracking, which routinely emerge from time to time~\cite{bounce-tracking-privacycg,bahrami2022fp,Sanchez2021JourneyToCookies,Munir23CookieGraphArxiv}.

Our key insight is that trackers need to share information (e.g., user/device identifiers) from the client to the server side regardless of what type of tracking is employed. 
Therefore, we contend that a promising orthogonal approach to anti-tracking is to detect and block the sharing of tracking information in network requests. 
Trackers commonly include tracking information in ``decorated'' network request URLs (aka link decoration).
Existing privacy-enhancing tools outright block network requests to endpoints using filter lists of known tracking services.
However, a decorated link can include both tracking (e.g., user/device identifiers) and functional (e.g., CSRF tokens, identifying news/product subpages) information, which renders existing request blocking countermeasures ineffective -- they risk breaking legitimate functionality as collateral damage if they block the request and risk allowing privacy-invasive tracking if they do not.
This is fundamentally a granularity issue. 
To the best of our knowledge, there is an unmet need for a fine-grained approach that can precisely remove tracking information in decorated links.

To sanitize link decoration (i.e., precisely removing just the tracking information from a request URL), privacy-focused browsers and browser extensions such as Safari, Firefox, Brave, uBlock Origin, and AdGuard employ a manually-curated list of query parameters that are known to be abused for tracking~\cite{privacytests,firefoxlistqueryparameters,brave2023filter,adguard-url-tracking-filter,ublockoriginlistqueryparameters}, similar to the request-blocking filter lists such as EasyList and EasyPrivacy \cite{easylist,easyprivacy}. 
While the manual approach might work for a small number of tracking query parameters, it cannot keep pace with the increasing adoption of query parameters by trackers \cite{randall2022measuring}, a problem that has also severely impacted filter lists~\cite{Alrizah19errorsfilterlists, Snyder20WhoFilterstheFiltersSigmetrics,Iqbal17AntiABIMC}.
Thus, it is not surprising that the filter lists to sanitize link decoration conservatively target only 10-100s of query parameters. 

In this paper, we propose \name (pronounced purel-l), a machine-learning approach to distinguish between tracking and functional link decorations.
\name makes use of a cross-layer graph representation to capture the complete execution of a webpage, which includes interactions between the HTML DOM structure, JavaScript execution behavior, information stored in browser storage, and network requests issued during a webpage load. 
\name then extracts distinguishing features from this rich graph representation and uses a supervised classifier to detect tracking link decorations. 

Our evaluation on a sample of top-million websites shows that \name can effectively (98.87\% recall) and safely (98.62\% precision) sanitize link decoration.
Overall \name achieves 98.74\% accuracy, significantly outperforming existing countermeasures by at least 7.71\% in terms of precision, 4.83\% in terms of recall, and 6.43\% overall accuracy while reducing website breakage by more than 8$\times$.
Our evaluation also shows that \name is robust against common evasion attempts such as changing link decoration names and splitting/combining link decoration values.

We deploy \name on a subset of top-million websites to measure the prevalence of link decorations for tracking.
\name detects that 73.02\% of the sites abuse link decorations for tracking, with an average site using 10.75 tracking link decorations. 
We find that the most common abusers of link decoration include well-known advertising and tracking services, which use link decorations to share cookies, email addresses, and fingerprinting information.

Our key contributions are as follows:
\begin{enumerate}

    \item We propose and evaluate \textbf{an automated machine learning approach}, called \name, to detect tracking link decorations using features that capture interactions and flow of information across multiple layers of the web stack.
    
    \item We \textbf{deploy \name on top-million sites} to measure the prevalence, abusers, and type of tracking information shared via link decorations.

    \item We  \textbf{use \name to generate a filter list}, which can and is already being used by privacy-focused browsers and browser extensions.
 
\end{enumerate}
\vspace{-10pt}

\section{Background \& Related Work}
\label{sec:background}
In this section, we discuss preliminaries, review related work, and survey existing countermeasures against the abuse of link decoration for tracking in industry and academia.

\subsection{What is Link Decoration?}
A URL is composed of the following key components: \textcolor{blue}{scheme}, \textcolor{green}{fully qualified domain name (FQDN)}, \textcolor{violet}{resource path}, \textcolor{red}{query parameters}, and \textcolor{orange}{fragments}.
We use \texttt{\textcolor{blue}{https}://\textcolor{green}{a.site.example}/\textcolor{violet}{YYY/ZZZ/}\textcolor{magenta}{pixel.jpg}?\\\textcolor{red}{ISBN=XXX\&UID=ABC123}\#\textcolor{orange}{xyz}}, as an example URL to define these segments below:

\para{Base URL}. \texttt{\textcolor{blue}{https://}} is the scheme in this URL and \texttt{\textcolor{green}{a.site.example}} is the FQDN. These segments are combined to form the base URL.

\para{Resource path}. Immediately following the FQDN, \texttt{\textcolor{violet}{/YYY/ZZZ}} is the resource path in the URL. 
It points to a directory, file, API endpoint, or other resource on the server. 
\texttt{\textcolor{magenta}{pixel.jpg}} is the name of the resource hosted on the server.

\para{Query parameters}. Immediately following the resource path after the $?$ delimiter, \texttt{\textcolor{red}{ISBN=XXX\&UID=abc123}} are the query parameters in the URL. 
A query parameter consists of a key-value pair, where the key is separated from the value by the \textit{=} delimiter.
Multiple query parameters in a URL are separated from each other by the $\&$ delimiter.

\para{Fragments}. Immediately following the query parameters after the $\#$ delimiter, \texttt{\textcolor{orange}{xyz}} is the fragment in the URL. 
Fragments can be a singular value or multiple key-value pairs that are separated by the \& delimiter (similar to query parameters).\footnote{In cases where fragments contain multiple key-value pairs, we treat these key-value pairs similarly as query parameters to facilitate easier comparison and analysis across different URL components.}

While only query parameters are traditionally considered as link decoration~\cite{Takata2021RiskAanalysisLinkDecoration, randall2022measuring}, we find that link decoration can be carried out using the resource path, query parameters, and fragments (discussed further in Section~\ref{subsec: methodology})\footnote{While fragments are not sent alongside a request to a server, the server can send a redirect to another page, which can access the fragments from the URL, e.g., through \texttt{window.location.hash}.}. Next, we describe the threat model we consider for tracking through link decoration.

\vspace{-10pt}
\subsection{Threat Model}
\label{app:threat_model}

Our threat model focuses on the sharing of identifying information through link decoration in third-party requests.
We exclude first-party requests from our analysis as we assume that first-parties would use identifying information to provide legitimate website functionality.
Our threat model considers the abuse of link decoration for both same-site and cross-site tracking because even same-site identifiers (e.g., first-party cookies \cite{Sanchez2021JourneyToCookies,ChenACM2021CookieSwapParty,Munir23CookieGraphArxiv}) can be combined with additional information for cross-site tracking. 
We consider two main entities in our threat model: the victim (users) and the adversary (third-party trackers).
While third-party trackers require some cooperation from the website publisher (i.e., embedding a script on the page), we do not consider the publisher to be an active part of the threat model.

{We assume that the victim/user:}

\noindent $\bullet$ has third-party and first-party cookies enabled in the browser

\noindent $\bullet$ may provide personally identifiable information (PII) such as email address on the website (e.g., to log in)

{We assume that the adversary/third-party tracker:}

\noindent $\bullet$ may be present in a third-party context or a first-party context (i.e., a script embedded in the main frame) on the websites visited by the user
    
\noindent $\bullet$ aims to collect identifying information such as identifier cookies and email addresses for tracking

As described below, third-party trackers can use link decorations to share identifying information in several ways:

A user visits a website \textit{website.example}, where tracker A is present in a first-party context (e.g., a script that sets a first-party identifier cookie) and tracker B is present in a third-party context (e.g., a pixel that sets a third-party identifier cookie).
The user provides their email address on a form field that is accessible to only tracker A.
To send the email address to its server, tracker A has to append the email address as a link decoration in the request URL to its server.
To send the first-party cookie to its server, tracker A again has to append the first-party cookie as a link decoration in the request URL to its server because first-party cookies would not be automatically sent to its third-party domain. 
Upon subsequent visits to the website by the same user, tracker A can associate its first-party cookie with the email address for same-site tracking even though the user may not provide the email address in subsequent visits.
Tracker A can also share the email address with tracker B by appending the email address as a link decoration in the request URL to tracker B's server.
Tracker B can cross-site track the user with its third-party cookie, while also being able to associate it with the email address shared by tracker A.

\vspace{-10pt}
\subsection{Abuse of Link Decorations for Tracking}
The abuse of link decoration for online tracking is not a new phenomenon. 
To the best of our knowledge, the earliest evidence of link decoration abuse is from 1996, when {Webtrends} (an analytics service) used the \texttt{WT.mc\_id} query parameter for click tracking in advertising campaigns \cite{webtrends1996, webtrends-documentation}.
Since then, link decorations in general, and query parameters specifically, have been widely used for creating personalized links to track the success of advertising campaigns.
For example, Urchin Tracking Module (UTM) parameters, such as \texttt{utm\_source}, are link decorations that identify the source of traffic on a website and attribute it to specific advertising campaigns~\cite{GoogleUTM, HubspotUTM, MailchimpUTM, MonsterinsightsUTM}.

Prior research has shown that trackers abuse link decoration to implement various tracking techniques \cite{Papadopoulos19cookiesyncing,Iqbal22USENIXKhaleesi,Fouad20PixelsPETS,PETS2021Article, Ren2021CNAME, Koop2020RedirectTracking,Munir23CookieGraphArxiv,DinoUSENIX22CookieBlock}. 
While prior work has proposed approaches to detect and block specific tracking techniques, which in turn rely on link decoration, these studies do not specifically study the abuse of link decorations for tracking.
To the best of our knowledge, Randall \etal \cite{randall2022measuring} is the first study to specifically study tracking query parameters. 
The authors found that 8.1\% of the navigation URLs are decorated with identifiers as query parameters for tracking.

With new restrictions \cite{firefox-cookie-policy, safari-default-cookie,Schuh2020ChromeCookies} on third-party cookies, trackers are moving towards alternative techniques of tracking, which include the use of first-party cookies \cite{Munir23CookieGraphArxiv,ChenACM2021CookieSwapParty, Sanchez2021JourneyToCookies}, personally identifiable information (PII) \cite{acar2020no, senol2022leaky_forms}, and device/browser fingerprinting \cite{Iqbal21FingerprintingSP, Fouad22CookieRespawningPETS, laperdrix2020browser}.
In contrast to third-party cookies that are automatically included as headers in outgoing HTTP requests, these alternative tracking techniques must rely on link decoration for sharing identifying information. 
As trackers shift their focus towards these alternative tracking techniques, it is reasonable to assume that the abuse of link decoration for tracking will also continue to increase.

\vspace{-10pt}
\subsection{Countermeasures Against the Abuse of Link Decorations for Tracking}
\label{subsec:countermeasures_related_work}
Given the increased focus on alternative techniques to track users due to restrictions on third-party cookies, privacy-focused browsers, and browser extensions have started deploying countermeasures against the abuse of link decorations. 
These countermeasures can largely be divided into two different categories: filter list based countermeasures that rely on a static filter list of link decorations and heuristic-based countermeasures that rely on certain properties to identify tracking link decorations.
Next, we describe existing countermeasures against tracking link decoration and highlight their limitations.  

\subsubsection{Filter List Based Countermeasures}

Filter lists of link decorations contain both site-specific and site-agnostic rules which determine which link decorations should be allowed and which link decorations should be removed.
These filter lists are manually curated and maintained, which has been shown by previous research to have issues such as slow updates and being error-prone \cite{Iqbal17AntiABIMC,hieu2021cv,alrizah2019errors}.
As discussed below, filter list based countermeasures are used by both privacy-focused browsers and browser extensions.

\para{Brave.} Since July 2020 \cite{brave2020grab}, Brave browser attempts to remove tracking query parameters from URLs by matching them against a list of known tracking query parameters.
This filter list is curated by analyzing the documentation provided by the trackers themselves and from the reports submitted by Brave developers and users.
At the time of writing, Brave's filter list of tracking query parameters contains 59 query parameters \cite{brave2023hacks,brave2023filter}. 
%


\para{Firefox.} In January 2022, Firefox introduced the {query parameter stripping} feature \cite{MozillaNightly2022Issue117} in Firefox Nightly 96. 

Mozilla integrated this feature in Firefox 102.0 in June 2022 \cite{mozilla2022firefox} although it was not enabled by default -- Firefox users have to set \textit{Strict} security level in {Enhanced Tracking Prevention} (ETP) to enable this feature.
Firefox also allows users to remove tracking link decorations from copied URLs \cite{Mozilla2023EnhancedTracking}.
Similar to Brave, Firefox also relies on a curated filter list of tracking query parameters. 
At the time of writing, Firefox's filter list of tracking query parameters contains 23 query parameters \cite{firefoxquerystripping}.

\para{Safari.} Since June 2023, Safari 17 \cite{safari17linkdecoration} removes known tracking query parameters in Safari's private browsing mode.
Similar to Brave and Firefox, Safari also relies on a curated filter list of tracking query parameters. 
At the time of writing, Safari's filter list of tracking query parameters contains 24 query parameters \cite{privacytests}.

\para{AdGuard.}
AdGuard introduced a new filter type named \texttt{removeparam} to remove tracking query parameters from request URLs in  2021 \cite{adguard-url-tracking-filter}.
At the time of writing, AdGuard's filter list includes more than one thousand query parameter rules \cite{adguard-url-tracking-filter-general,adguard-url-tracking-filter-specific,adguard-url-tracking-filter-allowlist}.

\para{uBlock Origin}. uBlock Origin introduced new filter types \texttt{queryprune} in 2020 and then switched to \texttt{removeparam} \cite{ublock-queryprune-github}.
Unlike AdGuard, uBlock Origin supports regular expression-based filters to remove tracking query parameters. 
At the time of writing, uBlock Origin includes 46 query parameter rules \cite{github-ublock-assets}.

\para{Requests based filter lists}. 
Filter lists such as EasyList \cite{easylist} and EasyPrivacy \cite{easyprivacy}, which are designed to block network requests to known trackers, would indirectly also block tracking link decorations.
However, blocking the whole URL is not practical where tracking and non-tracking link decorations are mixed in the same URL.
As we show later, using EasyList \cite{easylist} and EasyPrivacy \cite{easyprivacy} results in non-trivial false positives and false negatives.

\subsubsection{Heuristic Based Countermeasures}
The aforementioned filter list based countermeasures are limited because they need to be manually created and updated.
These limitations are apparent in their smaller size, with only one filter list including close to a thousand rules to detect tracking link decorations.

To address these issues, Randall \etal \cite{randall2022measuring} proposed CrumbCruncher -- a semi-automated heuristic-based approach to detect query parameters involved in the sharing of identifiers.
CrumbCruncher conducts parallel and consecutive crawls to identify query parameters that are distinct across parallel crawls but persistent across consecutive crawls.  
However, their approach is prone to false positives and false negatives. 
Specifically, not all persistent parameters are predisposed to be tracking, e.g., parameters such as \texttt{share\_button} or \texttt{en-US} remain consistent during multiple visits by the same user, however, they are not used for tracking.
The manual review by the authors showed that CrumbCruncher suffers from a 36\% false positive rate that needs to be addressed through a manual review.
%
%
%
%
CrumbCruncher also suffers from false negatives because it incorrectly assumes that tracking query parameters are persistent across subsequent crawls.
We find that almost 80\% of potential identifiers (i.e., longer than 8 characters as defined by CrumbCruncher) that are shared to known tracking endpoints (defined using EasyList \cite{easylist} and EasyPrivacy \cite{easyprivacy}) via query parameters change their value across consecutive crawls. 
For example, the \texttt{fbp} query parameter, which is used by Meta Pixel to share identifiers stored in the \texttt{\_fbp} cookie, does not maintain its value in about 87\% of the cases.\footnote{Bekos \etal \cite{bekos2023hitchhiker} showed that the value of the \texttt{\_fbp} cookie (and consequently the \texttt{fbp} query parameter) is randomly chosen from a list of up to 50 different identifiers, resulting in a new identifier for the same user each time.}
In summary, CrumbCruncher suffers from non-trivial false positives and false negatives due to its simplistic heuristic.
\vspace{-10pt}

\section{Motivating Measurements}
\label{sec:motivation}
In this section, we motivate the need for a tailored solution to curb the abuse of link decoration, such as \name, by demonstrating that link decoration abuse is a prevalent phenomenon.
Our key idea is to crawl a large set of websites and investigate the network requests, from known advertising and tracking services that appear on those websites, for link decoration. 
We exclusively focus on known advertising and tracking services because 
they are the main culprits who engage in such practices, and also because currently there are no current approaches to effectively detect link decoration abuse.

\vspace{-10pt}
\subsection{Methodology}
\label{subsec: methodology}

\para{Crawler configuration.}
We use OpenWPM (v0.17.0) \cite{Englehardt16OpenWPMCCS} and Firefox (v102) \cite{firefox102} for crawling.
Our crawls are stateless -- i.e., we clear all cookies and other local browser states before crawling each website.
By using stateless crawling, we ensure that each crawl is independent and not biased by the residual state from previous website crawls.
We turn off all built-in tracking protections provided by Firefox (Enhanced Tracking Protection [ETP]) \cite{firefox-etp}.
We conduct our crawls from the vantage point of an academic institution in the US.

\para{Websites crawled.}
We crawl a 20K sample of the Tranco top-million websites between March and April 2023
\cite{Pochat2019Tranco}.
As previous research has highlighted the importance of making crawls representative of sites with varying popularity \cite{worldwideview2022imc}, we crawl all of the top-1K sites, uniformly sample another 9K sites from the sites ranked 1K--100K, and a further 10K from sites ranked 100K--1M.
Additionally, to capture differing content on both the landing and internal pages \cite{aqeel2020landing}, we perform an interactive crawl that covers both types of pages.
Specifically, for each site, we crawl its landing page, randomly scroll and move the cursor (for bot mitigation), and then select up to 20 internal pages to visit at random.
After each page load is complete (i.e., when the onLoad event is fired), we uniformly at random wait an additional 5--30 seconds for bot mitigation and other resources to finish loading.
The success rate of our crawler is 98.79\%.
A tiny fraction of web pages do not load correctly because of server-side errors. 
\para{Labeling tracking requests.} 
To analyze the prevalence of link decoration in tracking requests, we use EasyList \cite{easylist} and EasyPrivacy \cite{easyprivacy}.
Specifically, we use them to label requests as Advertising and Tracking Service (\ats) or non-Advertising and Tracking Service (\nonats). 
We label a request as \ats if its URL matches the rules in either one of the lists. 
Otherwise, we label it as \nonats.

\para{Naming link decorations.}
\label{subsec:link-decoration-naming}
When link decorations are in the key-value format, the key can simply be combined with FQDN\footnote{We combine FQDN with the link decoration key because different FQDNs can use the same key names.} to uniquely identify a link decoration.
For example, if a link decoration with key \textit{username} is sent to an FQDN \textit{site.example.com}, \textit{site.example.com+username} can be used to identify the link decoration.
When link decorations are not in the key-value format (e.g., resource paths and fragments), we assign them keys based on the FQDN and their position in the URL.
We identify link decorations for resource paths based on their distance (directory levels) from the root. 
For the example URL: \texttt{\textcolor{blue}{https}://\textcolor{green}{a.site.example}/\textcolor{violet}{YYY/ZZZ/}\textcolor{magenta}{pixel.jpg}?\\\textcolor{red}{ISBN=ABC\&UID=DEF123}\#\textcolor{orange}{xyz}}, we identify the following link decorations as key-value pairs:

\begin{figure}[!t]
    \centering
    \includegraphics[width=\linewidth]{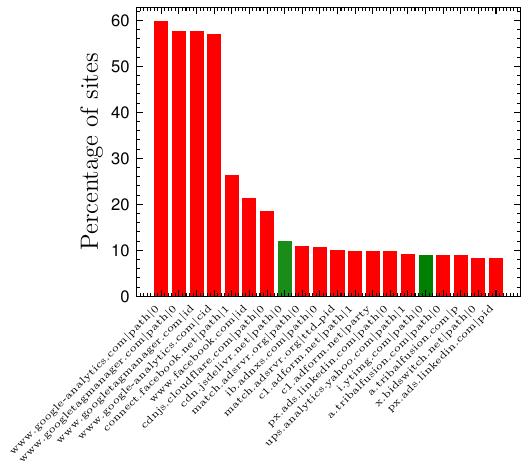}\
    \vspace{-25pt}
    \caption{Percentage of sites where the same link decoration by top domain appears and their primary usage. 
    The shades of red and green show link decoration's usage as \ats and \nonats, respectively.}
    \vspace{-10pt}
    \label{fig:link_decoration_prevalence}
\end{figure}

\noindent $\bullet$ \texttt{\textcolor{green}{a.site.example}$~|~ path_0$: \textcolor{violet}{YYY}}

\noindent $\bullet$ \texttt{\textcolor{green}{a.site.example}$~|~ path_1$: \textcolor{violet}{ZZZ}}

\noindent $\bullet$ \texttt{\textcolor{green}{a.site.example}$~|~ISBN$: \textcolor{red}{ABC}}

\noindent $\bullet$ \texttt{\textcolor{green}{a.site.example}$~|~UID$: \textcolor{red}{DEF123}}

\noindent $\bullet$ \texttt{\textcolor{green}{a.site.example}$~|~fragment$: \textcolor{orange}{xyz}}\footnote{If fragments are in the key-value format like query parameters, we treat them similarly as query parameters.}

This naming scheme allows us to compare link decoration values across different URLs.

\vspace{-10pt}
\subsection{Prevalence of Link Decoration}
\label{subsec:prevelance-link-decoration}
We investigate the prevalence of link decoration used on the 20K sample of the top-million websites.
Of the 44,648,436 link decorations in our data, 41.22\% are query parameters, 58.14\% are resource paths, and 0.63\% are fragments.
Considering only unique link decorations, we observe a total of 584,174 decorations: 42.41\% of which are query parameters, 53.85\% are resource paths, and 3.73\% are fragments.

Overall, 45.55\% unique link decorations are in the URLs labeled as \ats, while the rest are sent to \nonats endpoints.
We find that requests sent to \ats endpoints disproportionately contain more decorations on average (7.69) than the requests sent to \nonats endpoints (4.68), highlighting that link decorations are more frequently used by \ats than \nonats.
A similar trend holds for third-party requests (4.80) vs. first-party endpoints (2.10).

We further find that the same link decorations are widely reused (use of the same link decoration key/name on more than one site) by advertising and tracking services.
Figure \ref{fig:link_decoration_prevalence} shows the top 20 link decorations and their prevalence in our dataset.\footnote{To simplify the illustration due to space constraints, we limit link decorations for each FQDN to only the top two.}
The color of the bar shows the usage of link decoration for either \ats or \nonats purpose (i.e., the red represents \ats and green represents \nonats).
The plot shows that while both \ats and \nonats services show reuse of the same link decoration across multiple websites, it is \ats who predominately exhibit this behavior.
For example, \texttt{www.googletagmanager.com$|$id} is used in around 55\% of sites in our dataset and is primarily used in \ats requests.
On the other hand, the most commonly used \nonats link decoration is \texttt{cdnjs.cloudflare.com$|$path$|$0}, which is found on slightly more than 10\% of sites in our dataset.

\begin{figure}[!t]
    \centering
    \includegraphics[width=0.9\linewidth]{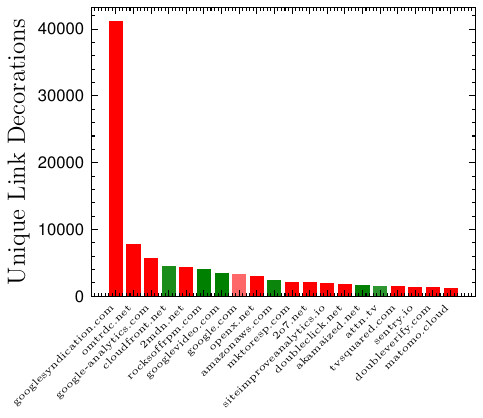}
    \vspace{-10pt}
    \caption{Total unique link decorations used by domains. The shades of red and green show link decoration's usage as \ats and \nonats, respectively.}
    \vspace{-10pt}
    \label{fig:link_decoration_count_domain}
\end{figure}

\begin{figure}[!t]
    \centering
    \includegraphics[width=0.9\linewidth]{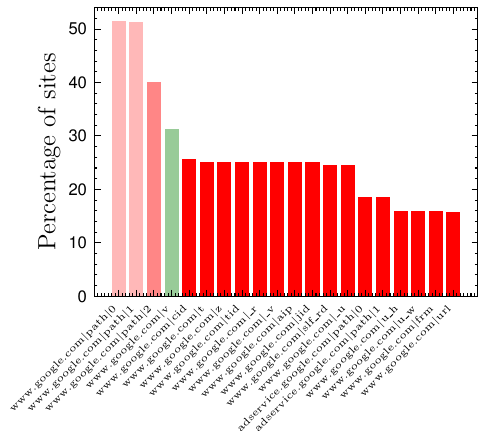}
    \vspace{-.2in}
    \caption{Average number of link decorations used by Google endpoints (minimum 1000 requests across 20K sites)}
    \label{fig:link_decoration_prevalence_single_domain}
    \vspace{-.2in}
\end{figure}

Next, we analyze the use of link decoration by \ats. 
Figure~\ref{fig:link_decoration_count_domain} plots top-20 tracking domains based on the number of unique link decorations they use.
The intensity of the color (green for \nonats and red for \ats) in the figure for each domain shows its use of link decorations in \ats or \nonats requests.
We note that \textit{googlesyndication.com}, which is used by Google Ad Manager \cite{googlesyndication-support-answer}, uses the highest number of unique link decorations among the \ats domains. 
It is followed by \textit{omtrdc.net}, which is used by Adobe Marketing Cloud \cite{adobe-marketing-cloud-omtrdc}, \textit{google-analytics.com}, which aggregates and reports user stats for sites \cite{google-analytics-documentation}, and \textit{cloudfront.net}, which is an Amazon-owned content delivery network \cite{amazon-cloudfront}.
Other well-known \ats such as Facebook, Baidu, and Microsoft are also among the domains that use the most link decorations in their requests.
Our main finding is that link decorations are widely used by well-known advertising and tracking services.
Crucially, Figure \ref{fig:link_decoration_count_domain} also shows some mixed usage of link decorations.
For example, link decorations used by \textit{google.com} were part of 122,028 \ats requests and 56,026 \nonats requests (this is represented by a lighter shade of red in the figure as compared to other domains which are either darker shades or red or green).
On the other hand, link decorations used by \textit{amazonaws.com} were part of 863 \ats requests and 16,420 \nonats requests (represented by a darker shade of green in Figure \ref{fig:link_decoration_count_domain}).

To identify the reason behind significant mixed usage of link decorations by \textit{google.com}, we take a closer look at different link decorations used by \textit{google.com} and their prevalence in our dataset.

Figure \ref{fig:link_decoration_prevalence_single_domain} plots the prevalence of the top 20 link decorations used by \textit{google.com}.
The color of each link decoration represents its use in \ats and \nonats link decorations, with higher shades of red representing predominant use in \ats requests and higher shades of green representing predominant use in \nonats requests.
We observe that the top 4 link decorations by \textit{google.com} have significant mixed usage between \ats and \nonats requests, with the top 3 link decorations: \texttt{www.google.com$|$path$|$0}, \texttt{www.google.com$|$path$|$1}, and \texttt{www.google.com$|$path$|$2} leaning towards more \ats use while \texttt{www.google.com$|$v} slightly leaning towards \nonats use.

\begin{figure}[!t]
    \centering
    \includegraphics[width=\linewidth]{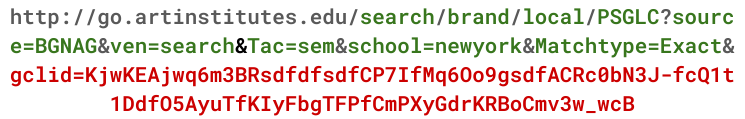}
    \vspace{-10pt}
    \caption{Example URL with mixed link decorations. \protect\tikz{\protect\draw[fill=black!40!green] rectangle(0.3, 0.2)} indicates a \nonats link decoration while \protect\tikz{\protect\draw[fill=black!30!red] rectangle(0.3, 0.2)} indicates an \ats link decoration. Resource paths are highlighted as green and the query parameters with keys \texttt{source}, \texttt{ven}, \texttt{Tac}, \texttt{school}, and \texttt{Matchtype} are used for functional purposes, while \texttt{gclid} contains an identifier that is used to track ad clicks.}
    \vspace{-10pt}
    \label{fig:mixed-url-example}
\end{figure}

These results show that a single link decoration can be part of both \ats and \nonats requests.
Next, we try to evaluate if a single request can also have both \ats and \nonats link decorations.
To this end, we make use of query parameter filter lists used by Brave\cite{brave2023filter}, Firefox\cite{firefoxlistqueryparameters}, and Safari\cite{privacytests}, as well as privacy-enhancing extensions uBlock Origin\cite{ublockoriginlistqueryparameters} and AdGuard\cite{adguard-url-tracking-filter-general, adguard-url-tracking-filter-specific}.
We label every link decoration not included in these filter lists as a \nonats link decoration.
Overall, we observe 51,736 requests that contain one or more \ats link decoration, while only 248 of these requests contain no other \nonats link decoration.
On average, an \ats link decoration is accompanied by 16.06 \nonats link decorations in the same request URL.
An example of such mixed URLs is shown in Figure \ref{fig:mixed-url-example}.

\para{Takeaway.}
Our measurements show that a request URL can contain both \ats and \nonats link decorations.
Moreover, the classification of a link decoration can change depending on which site it is used on and the domain it is being sent to.
Thus, as we demonstrate later in Section \ref{subsec:comparison}, it is not trivial to detect and block \ats link decorations using existing countermeasures.
\vspace{-10pt}

\section{\name}
\label{sec:counter_measures}
In this section, we present \name (pronounced purel-l), our machine learning approach to detect \ats link decorations.
\name's key idea is to use the execution traces of \ats link decorations as their signatures, which it learns and automatically detects with the help of a machine learning (ML) classifier.
\name captures detailed execution traces across the HTML, network, JavaScript, and storage layers of the web stack and models them in a graph representation. 
The graph representation captures the natural interaction between different layers of the web stack and provides a parse-able representation to extract various characteristics (i.e., features) of \ats link decoration execution, that are used to train a supervised ML classifier. 
Figure \ref{fig:decograph-methodology} provides an overview of \name's design.

\begin{figure*}[!t]
    \centering
    \includegraphics[width=\linewidth]{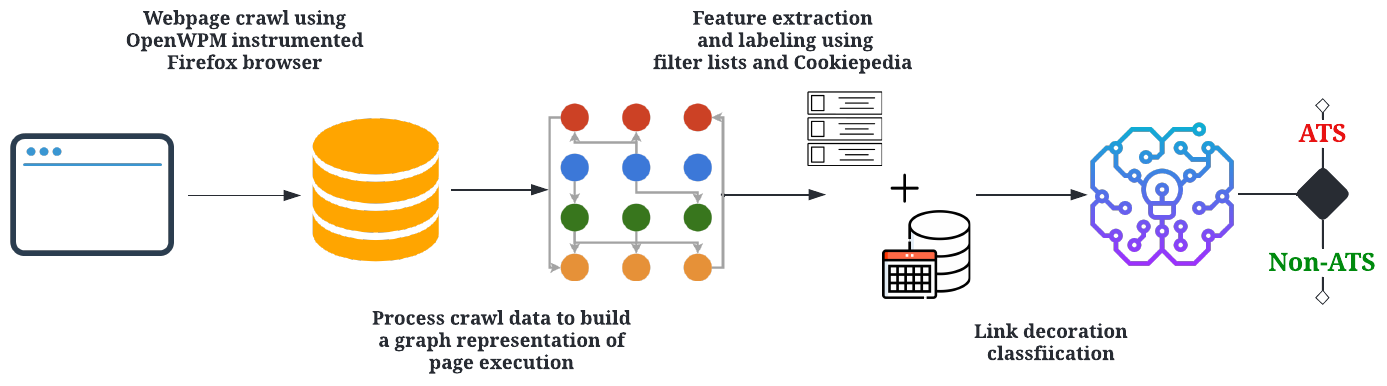}
    \vspace{-10pt}
    \caption{Overview of \name pipeline: (1) Webpage crawl using an instrumented browser; (2) Construction of a graph representation to represent the instrumented webpage execution information; (3) Feature extraction for graph nodes that represent link decorations; and (4) Classifier training to separate out \ats and \nonats link decorations.}
    \vspace{-10pt}
    \label{fig:decograph-methodology}
\end{figure*}

\vspace{-10pt}
\subsection{Design and Implementation}
\label{subsec:design}
\para{Browser instrumentation.} \name extends OpenWPM \cite{Englehardt16OpenWPMCCS}, an open-source web measurement tool, to record the execution of a webpage across \textit{HTML}, \textit{network}, \textit{JavaScript}, and \textit{storage} layers during a webpage load. 
Similar to prior work on tracking detection (e.g.,  \cite{Iqbal20AdgraphSP,Siby22WebGraph, Munir23CookieGraphArxiv}), \name captures the HTML elements created by scripts, network requests sent by scripts and HTML elements, responses received to these requests, sharing of identifiers stored in storage (local storage and cookies), and other read/write operations on storage mechanisms present in the browser.

\name improves upon previous work by creating a more granular representation of the network layer of a webpage, which is essential for capturing characteristics of link decoration.
Specifically, instead of coarsely capturing network requests and responses, \name breaks them down and captures granular components of link decorations.
For example, instead of identifying that a request contains a cookie value, PURL identifies the exact link decoration that was used to share that cookie value.

\para{Graph Construction.}
There are five types of nodes in \name's graph representation: \textit{storage}, \textit{HTML}, \textit{script}, \textit{network}, and \textit{decoration}.
\textit{Storage} nodes refer to information stored in cookies and localStorage.
\textit{HTML} nodes refer to HTML Document Object Model (DOM) elements on a webpage.
\textit{Script} nodes map interactions of JavaScript execution on a webpage.
\textit{Network} nodes represent outgoing HTTP requests and incoming HTTP responses from network endpoints.
\textit{Decoration} nodes are created by splitting each \textit{network} node into its link decorations.

We capture read and write operations performed on information in browser storage by different scripts, their sharing through network requests, and the setting of browser storage through network responses.
We capture the interaction of different scripts with HTML elements and also map requests generated through HTML elements.
In addition to these interactions, \name captures actions and attributes that are specific to link decorations.
Since link decorations may be used to share information stored in browser storage and the network responses received as a result of this sharing may be used to set browser storage elements using HTTP headers or JavaScript APIs, we monitor Base64-, MD5-, SHA-1-, and SHA-256-encoded\footnote{\name can be extended to support other encodings.} storage node values in decorations to associate relevant interactions between decoration, storage, network, and script nodes.

Figure \ref{fig:decograph-example} and \ref{fig:decograph-representation} show how \name constructs a graph representation for an example scenario where a script is reading/writing to browser storage and sending requests including link decorations to tracking sources.
The nodes in the given example graph are storage, script, request, and decoration.
The numbers on the edges represent a particular action, as represented in Figure \ref{fig:decograph-example}.
Dotted and dashed lines respectively show the flow of information from storage to decoration nodes (exfiltration) and the flow of information from request nodes to storage nodes (infiltration).
\name links the outward flow of information (exfiltration) to decoration nodes, and also maps the inward flow of information (infiltration) from a parent request/response node to the storage node.
We use this graph structure to calculate features for decoration nodes that represent this flow of information.

\begin{figure*}[!t]
    \centering
    \includegraphics[width=\linewidth]{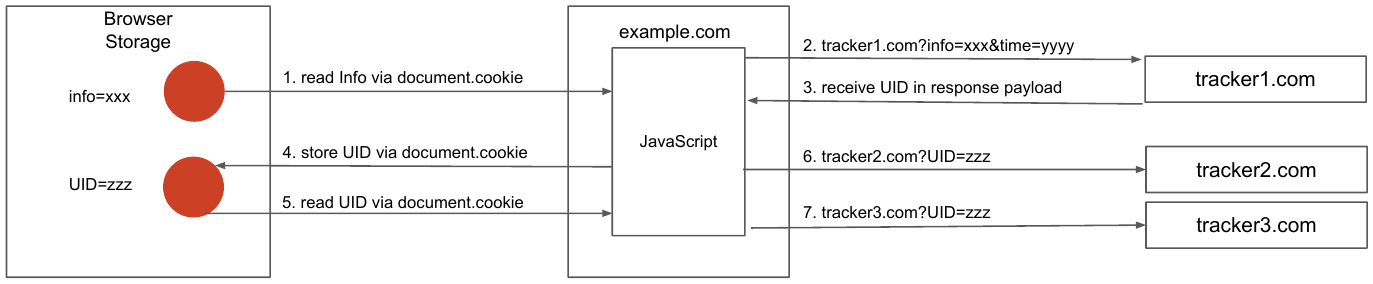}
    \vspace{-10pt}
    \caption{Example scenario to illustrate \name's graph construction (shown in Figure~\ref{fig:decograph-representation}). (1) A script on example.com reads info cookie from browser storage using \texttt{document.cookie}. (2) The script sends a network request to \textit{tracker1.com} which includes the info cookie value and the current time in the decorated link. (3) \textit{tracker1.com} sends a network response that contains UID in the response payload. (4) The script stores UID in the browser storage using \texttt{document.cookie}. (5) The script reads UID  using \texttt{document.cookie} (6,7) The script sends the UID to \textit{tracker2.com} and \textit{tracker3.com} as decorated links.}
    \vspace{-10pt}
    \label{fig:decograph-example}
\end{figure*}

\begin{figure}[!t]
    \centering
    \includegraphics[width=.9\linewidth ]{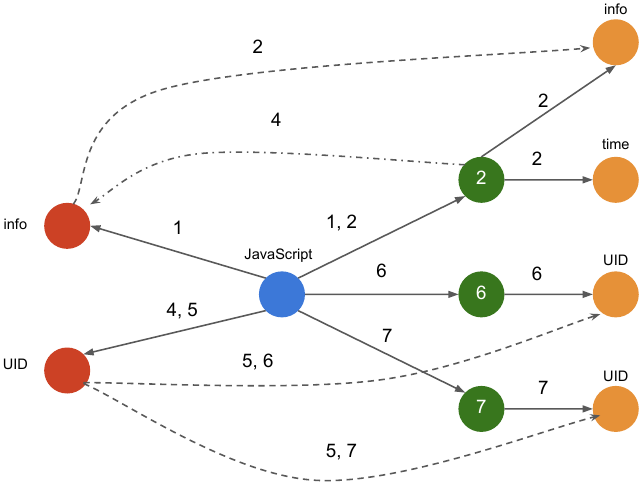}
    \vspace{-10pt}
    \caption{Graph representation of Figure~\ref{fig:decograph-example} in \name. \protect\tikz{\protect\draw[fill=black!40!green] circle(1.0ex);} network nodes, \protect\tikz{\protect\draw[fill=black!0!blue] circle(1.0ex);} script nodes, \protect\tikz{\protect\draw[fill=black!-10!red] circle(1.0ex);} storage nodes, and
    \protect\tikz{\protect\draw[fill=black!-40!orange] circle(1.0ex);} decoration nodes. While the solid lines show the interactions of the script nodes with the storage and request nodes, the dashed (- - -) and dotted (. \_ .) lines represent the flow edges that are captured by \name.}
    \vspace{-10pt}
    \label{fig:decograph-representation}
\end{figure}

\para{Feature extraction. }
\name leverages the graph representation to extract three different types of features that capture the execution traces of link decoration, referred to as \textit{structural}, \textit{flow}, and \textit{content} features.

\textit{Structural} features map the relationship of different nodes in the graph with each other, such as the connectivity of nodes and information about their ancestry.
For example, the connectivity of nodes can capture how many different link decorations are in a request.
As described in Section \ref{sec:motivation}, on average, \atses use far more link decorations than \nonatses, resulting in stronger connectivity for \ats link decorations that will be reflected in corresponding structural feature values.
Structural features help encode this information for our classifier using graph properties such as centrality, connectivity, and closeness\cite{Maurya2021GraphNeural, Roy2014ModelingMeasuring}.

\textit{Flow} features represent information flow across different layers of a webpage.
Capturing this flow of information is important to track how user information is, extracted, stored, and sent out through tracking link decorations.
First, \name captures the direct sharing of storage values through the link decoration.
As described in Figure \ref{fig:decograph-example}, storage values can directly be sent out using link decorations.
\name represents this sharing as additional edges from storage nodes to decoration nodes.
Second, \name also keeps track of whether the parent request of a link decoration results in the setting of a storage node.
This inward flow of information (infiltration) is usually used by \ats to set identifiers based on information sent out in requests (mainly through link decorations) \cite{Munir23CookieGraphArxiv}.
\name maps this infiltration through indirect edges connecting the parent request/response of decoration with the corresponding storage node.
In addition, \name monitors if the script sending the parent request of decoration is involved in sending storage information in non-parent requests or is part of redirects \cite{Iqbal22USENIXKhaleesi}.

To determine the suitability of a decoration as a potential identifier, \name also computes \textit{content} features such as character-level Shannon entropy \cite{shannon1948entropy} and the relative position of the decoration in the URL.
The complete list of the features used by \name and their analysis is in the appendix.

\para{Ground Truth Labeling. }
Once we capture the execution traces, we need to label them before they can be used to train a classifier. 
However, as discussed in Section~\ref{sec:background}, there are currently no readily available sources that can be reliably used to label link decoration abuse. 
Thus, we create our own set of labels by combining three different sources, which are: (i) filter lists of known advertising and tracking sources, (ii) a database of known tracking cookies, and (iii) short lists of manually curated tracking query parameters.
Recall from Section~\ref{subsec:link-decoration-naming} that each link decoration instance is a combination of the site where it appeared and the decoration key, and it is labeled as such. 
Next, we describe our ground truth labeling process that leverages the three aforementioned approaches.

\textit{1) Filter lists.} 
Filter lists, such as EasyList and EasyPrivacy \cite{easylist, easyprivacy}, are the most reliable sources to identify tracking, which are used by almost all privacy-enhancing tools. 
However, their detection granularity is at the level of a URL and thus cannot be directly used to label individual parameters as \ats in a URL. 
This is because even a URL detected as \ats by filter lists might contain both tracking and non-tracking parameters~\cite{brave-github-query-filtering, firefox-doc-link-decoration}. 
Despite this problem, filter lists can still be used to identify benign parameters, i.e., the parameters found in URLs from \nonats services, and we use them as such.  
Specifically, we rely on EasyList \cite{easylist} and EasyPrivacy \cite{easyprivacy} filter lists to first identify \nonats URLs and then label all parameters in them as \nonats.
As shown in Section \ref{subsec:prevelance-link-decoration}, a single URL can contain both \ats and \nonats link decorations.
In such cases, labeling all link decorations in URLs that are not blocked by filter lists as \nonats will result in incorrect labels for \ats link decorations.
To account for this, we re-label all \nonats link decorations identified in this step as \ats if they are found to be involved in tracking in the next steps.

\textit{2) Cookiepedia.}
Link decoration can contain values stored in cookies, which are traditionally used to store and share user identifiers \cite{Sanchez2021JourneyToCookies, Munir23CookieGraphArxiv}.
To identify such link decorations that can be used to exfiltrate tracking cookies, we make use of Cookiepedia \cite{cookiepedia}, which is a database of cookies maintained by a well-known Consent Management Platform (CMP) called OneTrust \cite{hils2020measuringcmp, DinoUSENIX22CookieBlock}.
Primarily, Cookiepedia provides the purpose of each cookie in its database through its integration with OneTrust.
Each cookie is provided one of the four labels: strictly necessary, functional, analytics, and advertising/tracking.
We monitor the sharing of all cookies labeled as either analytics or advertising/tracking in Cookiepedia through a link decoration and label those link decorations as \ats. 

\textit{3) Manually Curated Lists.}
Recall from Section~\ref{sec:background} that several privacy-enhancing browsers and extensions maintain lists of known \ats link decorations parameters. 
Despite these lists being limited, they contain popular query parameters that are manually vetted to be used for tracking. 
Thus, we also make use of \ats link decoration lists, maintained by Brave \cite{bravelistqueryparameters}, Firefox \cite{firefoxlistqueryparameters}, AdGuard \cite{adguardlistqueryparameters}, and uBlock Origin \cite{ublockoriginlistqueryparameters}.

Using a combination of these techniques, we were able to label 18.76\% of our dataset, with 1.40\% (60,573 instances) being labeled as \ats and 17.36\% (749,553) as \nonats.
Even though the ground truth is limited, especially for \ats samples, we argue that it is a significant improvement over the existing countermeasures.
As described in Section \ref{subsec:countermeasures_related_work}, prior work has identified only a handful (maximum of around one thousand) \ats link decorations.
Additionally, mislabelling link decorations can result in significant website breakage, necessitating a high-precision, albeit limited, ground truth.

\para{Classifier.}
After curating the ground truth, we next train a supervised classifier to detect \ats link decoration. 
%
We use a random forest ensemble classifier because it is tolerant against noisy labeled data, is efficient to train, and is interpretable \cite{uddin2022leveraging, mills2019efficient}.
We train the model using a balanced set of \ats and \nonats link decoration samples. 
We evaluate the accuracy of our classifier using stratified 10-fold cross-validation, to ensure that we do not train and test on the same samples.
Overall, our classifier achieves 98.74\% accuracy, 98.62\% precision, and 98.87\% recall, indicating that it is successful in detecting \ats link decorations.

The classifier accuracy is also comparable across different types of link decorations.
For resource paths, it achieves an accuracy of 99.36\%, 99.03\% precision, and 99.69\% recall. For query parameters, the accuracy is 96.40\%, precision is 93.39\%, and recall is 99.87\%. Finally, for fragments the accuracy is 99.33\%, precision is 98.75\%, and recall is 100\%.

\vspace{-10pt}
\subsection{Analysis of Disagreements between \name and Ground Truth}
\vspace{-5pt}
We manually analyze \name's false positives and false negatives to assess whether these are actual mistakes or limitations of our ground truth (recall that we curated a high-precision, albeit limited, ground truth).

First, we analyze the false positives of \name. 
We manually verify the most common false positives in our dataset by first, analyzing information sent by false positive link decorations (user identifiers, values stored in cookies, etc.), and second, by analyzing available online documentation by senders/receivers of these link decorations.
Our analysis of the most commonly misclassified \ats link decorations reveals that most of them are indeed used for tracking.
In total, \name classifies 8,058 \nonats instances (out of 749,553 total, 1.07\%) as \ats, which correspond to 2,994 unique link decorations.
The three most common false positive link decorations include \texttt{utk} query parameter sent to \textit{hubspot.com}, \texttt{bsi} query parameter sent to \textit{frog.wix.com}, and \texttt{iiqpciddate} query parameter sent to \textit{api.intentiq.com}.
We manually analyzed these three link decorations, which account for almost 10\% of all false positives.
Our analysis of the documentation for these three link decorations shows that these are not false positives, but rather these were falsely labeled as \nonats in our ground truth, and \name actually correctly classified them as \ats.
\texttt{utk} query parameter contains the HubSpot's \texttt{hubspotutk} cookie which is used to identify a user visiting a website \cite{utkfalsepositive}; \texttt{bsi} query parameter contains the identifier used by BSI's Customer Data Platform \cite{bsifalsepositive}, while \texttt{iiqpciddate} accompanies \texttt{iiqpcid} query parameter which is used by IntentIQ to uniquely identify a user \cite{intentiqfalsepositive}.
We conclude that \name's false positives are, in actuality, false negatives in the ground truth, which was curated conservatively to be highly precise (rather than high recall).

Second, we analyze the false negatives of \name.
In total, \name incorrectly classifies only 50 \ats instances as negatives (out of 60,573 total, 0.08\%).
In all of these instances, \name was unable to link the sharing of stored information through these link decorations.
In addition to this, values of structural features (e.g., number of edges) for these instances were also lower than true positive instances (e.g. 11,501.99 number of edges as compared to 14,762.86 for true positives).
We conclude that false negatives happen when \name is unable to trace certain tracking behaviors of \ats link decorations.
We elaborate on \name's implementation limitations in tracing storage sharing in Section \ref{sec:limitations}.

Beyond these disagreements, \name is more than the sum of its ground truth. 
Concretely, \name detects 52,489 \ats link decorations that are not detected by Cookiepedia, EasyList \cite{easylist}, EasyPrivacy \cite{easyprivacy}, or the manually curated filter lists for \ats link decorations.
\vspace{-10pt}

\vspace{-5pt}
\subsection{Comparison with Existing Countermeasures}
\label{subsec:comparison}
\vspace{-5pt}
In this section, we compare \name versus existing countermeasures against \ats link decorations to demonstrate that it significantly advances the state-of-the-art.
We compare it against approaches that directly detect \ats link decoration and also approaches that were originally designed to detect \ats link decoration but can be repurposed to detect them.

We compare \name against CrumbCruncher~\cite{randall2022measuring} and link decoration based filter lists, which are designed to detect \ats link decoration.
For comparison with CrumbCruncher, we rely on the list of \ats query parameters published by Randall \etal \cite{crumbcruncherlistqueryparameters}.
For comparison with link decoration based filter lists, we rely on the lists offered by Brave \cite{bravelistqueryparameters}, Firefox \cite{firefoxlistqueryparameters}, Safari \cite{privacytests}, uBlock Origin \cite{ublockoriginlistqueryparameters}, and AdGuard \cite{adguardlistqueryparameters}).

Additionally, we also compare \name against 
Cookiepedia~\cite{cookiepedia} and request-based filter lists (i.e., EasyList \cite{easylist} and EasyPrivacy \cite{easyprivacy}), which we repurpose to detect \ats link decoration.
We compare with Cookiepedia cookie labels by identifying link decorations that are used to exfiltrate values of cookies labeled as \ats by Cookiepedia.
For comparison with request-based filter lists, we consider all link decorations in the requests labeled as \ats by filter lists to be \ats link decorations.
We compare these approaches across two axes: (i) accuracy and (ii) breakage.

\para{Accuracy.}
Table \ref{tab:purl-comparison} shows the comparison of accuracy, precision, and recall of all four countermeasures against \name.
It can be seen from the table that \name outperforms the runner-up countermeasure (i.e., request-based filter lists) by 6.43\% in terms of accuracy and by 7.71\% in terms of precision.
Some of the most common \ats link decorations detected by \name and missed by the runner-up countermeasure are \texttt{sync.intentiq.com$|$pcid}, \texttt{pr-bh.ybp.yahoo.com$|$path$|$2}, and \texttt{partner.mediawallahscript.com$|$uid}. 
As we discuss next, the higher precision results in a measurable reduction in website breakage when \name is used as compared to runner-up countermeasures.

\begin{table}[!t]
  \centering
    \resizebox*{\linewidth}{!}{%
  \begin{tabular}[c]{ l c c c}
    \toprule
    \textbf{Classifier}                 & \textbf{Accuracy} & \textbf{Precision} & \textbf{Recall} \\
    \midrule
    \textbf{\name}                      & \accuracy\%       & \precision\%       & \recall\%       \\
    \textbf{CrumbCruncher}              & 50.16\%           & 59.09\%            & 10.67\%         \\
    \textbf{Cookiepedia}                & 80.99\%           & 99.01\%            & 62.63\%         \\
    \textbf{Filter lists (Requests)}    & 92.31\%           & 90.91\%            & 94.04\%         \\

    \textbf{Filter lists (Decorations)} & 50.50\%           & 100.0\%            & 10.15\%         \\

    \bottomrule
  \end{tabular}
    }
  \vspace{-10pt}
  \caption{Classification accuracy of \name, CrumbCruncher, Request Filter lists, Decoration Filter lists, and Cookiepedia}
  \vspace{-10pt}
  \label{tab:purl-comparison}
\end{table}
\begin{table}[!b]
\vspace{-.2in}
  \centering
  \resizebox*{\linewidth}{!}{%
    \begin{tabular}{l  c  c  c  c  c  c  c  c}
      \toprule
      \multirow{2}{*}{\textbf{Classifier}}    & \multicolumn{2}{c}{\textbf{Navigation}} &
      \multicolumn{2}{c}{\textbf{SSO}}        &
      \multicolumn{2}{c}{\textbf{Appearance}} &
      \multicolumn{2}{c}{\textbf{Miscellaneous}}                                                                                                                                                                                                          \\
                                              & Minor                                   & Major                & Minor                & Major                & Minor                & Major                & Minor                 & Major                \\
      \midrule
      \name                                   & \cellcolor{green}0\%                   & \cellcolor{green}0\% & \cellcolor{orange}2\% & \cellcolor{green}0\% & \cellcolor{orange}2\% & \cellcolor{red}2\% & \cellcolor{green}0\%  & \cellcolor{green}0\% \\

      Filter lists                             & \cellcolor{green}0\%                    & \cellcolor{red}2\% &
      \cellcolor{orange}2\%                    & \cellcolor{green}0\%                      & \cellcolor{orange}8\% &
      \cellcolor{red}6\%                      & \cellcolor{orange}4\%                    & \cellcolor{green}0\%   \\       
      \bottomrule
    \end{tabular}}
    \caption[]{Website breakage comparison of all three countermeasures.(\greenline) signifies no breakage, (\orangeline) minor breakage, and (\redline) major breakage. Each cell represents the percentage of sites on which breakage was observed.
  }
  \label{tab:breakge_overview}
\end{table}

\para{Website Breakage.}
To determine what effect each countermeasure has on the usability of a site, we compare website breakage caused by PURL and the countermeasure with the second-highest accuracy and recall (request-based filter lists) on 50 sites.
We sample 25 sites sampled from 20K sites used in section \ref{sec:measurements}, out of which 25 sites were sampled from those ranked 1-1000 and 25 from the rest to ensure the sample is representative.
A list of sites used for breakage analysis is available at \cite{PURLSanitizerBreakageAnalysis}.
Our breakage analysis is divided into four different categories of how a website is used: navigation (moving from one page of the website to another), SSO (third-party login integrations), appearance (visual consistency and acuity), and miscellaneous (additional functionality such as shopping carts, chatbots, etc.).
We also categorize each breakage into minor or major, with the former implying that the underlying functionality is disrupted but still usable and the latter implying that the functionality is completely unusable from the user's perspective.
Two reviewers interact with the webpage while \ats link decorations detected by \name or filter lists are removed.
Any disagreements between the two reviewers are resolved after careful discussion.

Our evaluation shows that out of the 50 sites, \name caused minor breakage on 2 sites, while it caused major breakage due to a failure to load CSS on \textit{autodesk.com}. On the other hand, requests-based filter lists caused minor breakage on 7 different sites and major breakage on 4 sites, including a complete breakdown of navigation on \textit{directunlocks.com} and CSS issues on \textit{engadget.com} and \textit{mysuncoast.com}.

These results show that \name not only significantly outperforms existing countermeasures in terms of accurately sanitizing more \ats link decorations, but it also does so without causing additional website breakage.
\vspace{-5pt}

\section{Deployment}
\label{sec:measurements}
In this section, we deploy \name on a 20K sample of top-million sites to understand the prevalence and the nature of information shared in \ats link decorations.

\vspace{-5pt}
\subsection{Prevalence of \ats Link Decorations}
We first analyze the breakdown of \name's classification of link decorations.
\name classifies 4.56\% (196,890) of link decorations in our dataset as \ats.
73.02\% (14,604) of the tested sites contain at least one request with an \ats link decoration.
Overall, an average site employs 10.75 \ats and 44.59 \nonats link decorations.
Table \ref{tab:link-decoration-stats-overall} provides the breakdown of different types of \ats and \nonats link decorations.
Out of the 196,890 link decorations labeled as \ats, 6.62\% are resource paths, 93.36\% are query parameters, and 0.02\% are fragments.
Our findings indicate that while query parameters account for the majority of \ats link decorations, trackers also abuse resource paths and fragments that are ignored in prior work \cite{randall2022measuring}.

\begin{table}[!t]
    \centering    
    \resizebox{\linewidth}{!}{
    \begin{tabular}{l|c c c|c}
        \toprule
        {} & \textbf{Resource Paths} & \textbf{Query Parameters} & \textbf{Fragments} & \textbf{Total}\\
        \midrule
        \textbf{\ats} & 13,030 & 183,813 & 47  & 196,890 \\
         & (6.62\%) & (93.36\%) & (0.02\%) & (100.00\%) \\
    \hline
        \textbf{\nonats} & 1,295,246  & 2,824,549 & 2,233  & 4,122,028 \\
         & (31.42\%) &  (68.52\%) & (0.05\%) & (100.00\%)  \\

        \midrule
        \textbf{Total} & 1,308,276  & 3,008,362  & 2,280  & 4,318,918 \\
         & (30.29\%) & (69.65\%) & (0.05\%) &  (100.00\%)\\
        
        \bottomrule
    \end{tabular}
    }
    \vspace{-10pt}
    \caption{Distribution of link decorations between \ats and \nonats across the 20K sample of top-million sites}
    \vspace{-10pt}
    \label{tab:link-decoration-stats-overall}
\end{table}

Table \ref{tab:top_50_combined_link_decorations_cookies} lists the top-50 most prevalent \ats link decorations.
We note that a majority of the top \ats link decorations are used by various Google advertising and tracking endpoints on \textit{doubleclick.net}, \textit{google-analytics.com}, and \textit{google.com}. 
For example, \texttt{cid} and \texttt{\_gid} are used by all of the aforementioned Google domains, with \texttt{cid} and \texttt{\_gid} on \textit{www.google-analytics.com} shared on more than half of the sites. 
After Google, \texttt{fbp} on \textit{www.facebook.com} used by Meta pixel \cite{fbp-fbc-parameters} is present on nearly 20\% of the sites. 
After Facebook, \texttt{vid} and  \texttt{sid} on \textit{bat.bing.com} used by Microsoft/Bing Ads Conversion tracking \cite{MicrosoftConversionTracking} are present on about 5\% of the sites. 
All of the remaining top-50 \ats link decorations also belong to well-known advertising and tracking organizations such as BidSwitch \cite{BidSwitchUserMatching}, LiveRamp \cite{LiveRampCookieSyncTag}, Yahoo! \cite{DuckDuckGoTrackerRadarYahoo}, Magnite \cite{DuckDuckGoTrackerRadarRubiconProject}, Amazon \cite{DuckDuckGoTrackerRadarAmazonAdSystem}, Criteo \cite{DuckDuckGoTrackerRadarCriteo}, OpenX \cite{DuckDuckGoTrackerRadarOpenX}, Oracle/BlueKai \cite{DuckDuckGoTrackerRadarBlueKai}, The Trade Desk \cite{DuckDuckGoTrackerRadarAdsrvr}, Sovrn Holdings \cite{DuckDuckGoTrackerRadarLijit}, Index Exchange \cite{DuckDuckGoTrackerRadarCasaleMedia}, TripleLift \cite{DuckDuckGoTrackerRadar3Lift}, and LiveIntent \cite{DuckDuckGoTrackerRadarLiadm}.
The presence of these advertising and tracking organizations in top-50 \ats link decorations (and even beyond top-50, not shown here due to space constraints) highlights \name's effectiveness in detecting both popular and relatively lesser known \ats link decorations.

Next, we investigate different types of information shared through these \ats link decorations.

\begin{table}[!t]
    \centering
  	\resizebox{\linewidth}{!}{
   \begin{tabular}{c|l c r}
\toprule
         \textbf{FQDN} &                         \textbf{Key} &  \textbf{Storage } &  \textbf{\%  } \\
        \textbf{} &                         \textbf{} &  \textbf{Keys} &  \textbf{Sites} \\
        
\midrule
          www.google-analytics.com &                          cid &      251 &             56.84 \\
      www.google-analytics.com &                         \_gid &      206 &             50.90 \\
       stats.g.doubleclick.net &                          cid &      220 &             36.60 \\
       stats.g.doubleclick.net &                         \_gid &      133 &             34.04 \\
                www.google.com &                          cid &      193 &             23.43 \\
              www.facebook.com &                          fbp &        5 &             19.14 \\
   googleads.g.doubleclick.net &                         auid &       46 &             13.41 \\
          analytics.google.com &                          cid &       98 &             11.03 \\
   googleads.g.doubleclick.net &                       cookie &        1 &              7.72 \\
   googleads.g.doubleclick.net &                         gpic &        1 &              7.71 \\
   googleads.g.doubleclick.net &                       ga\_vid &      112 &              6.97 \\
securepubads.g.doubleclick.net &                       ga\_vid &       54 &              6.69 \\
  partner.googleadservices.com &                       cookie &        1 &              5.60 \\
  partner.googleadservices.com &                         gpic &        1 &              5.59 \\
                  bat.bing.com &                          vid &        2 &              5.55 \\
                  bat.bing.com &                          sid &        2 &              5.54 \\
securepubads.g.doubleclick.net &                       cookie &        1 &              5.45 \\
securepubads.g.doubleclick.net &                         gpic &        1 &              5.42 \\
      www.google-analytics.com &                          sid &      802 &              4.99 \\
               x.bidswitch.net &                      user\_id &      139 &              4.70 \\
              idsync.rlcdn.com &                  partner\_uid &      208 &              4.61 \\
               pixel.tapad.com &            partner\_device\_id &      174 &              4.50 \\
          cm.g.doubleclick.net &                    google\_hm &       77 &              4.28 \\
       ups.analytics.yahoo.com &                          uid &      182 &              4.19 \\
      pixel.rubiconproject.com &                          put &      185 &              3.96 \\
              www.facebook.com &                           ts &      202 &              3.84 \\
securepubads.g.doubleclick.net &                       ga\_sid &      618 &              3.79 \\
 pagead2.googlesyndication.com &                          rst &      115 &              3.50 \\
              www.facebook.com &                           it &      167 &              3.48 \\
                www.google.com &                         auid &       29 &              3.48 \\
         s.amazon-adsystem.com &                           id &      213 &              3.45 \\
                gum.criteo.com &                         info &        2 &              3.30 \\
                us-u.openx.net &                          val &      132 &              3.14 \\
           pr-bh.ybp.yahoo.com &                        path | 2 &       47 &              3.13 \\
securepubads.g.doubleclick.net &                           dt &      117 &              3.12 \\
              tags.bluekai.com &                           id &       90 &              3.07 \\
   googleads.g.doubleclick.net &                          fst &      188 &              3.05 \\
   googleads.g.doubleclick.net &                       ga\_sid &      741 &              2.95 \\
              match.adsrvr.org &                     ttd\_puid &       17 &              2.92 \\
   googleads.g.doubleclick.net &                       random &      182 &              2.91 \\
                  ce.lijit.com &                         3pid &      167 &              2.89 \\
                www.google.com &                       random &      182 &              2.85 \\
      dsum-sec.casalemedia.com &             external\_user\_id &      126 &              2.85 \\
securepubads.g.doubleclick.net &                          lmt &      414 &              2.77 \\
                 eb2.3lift.com &                         xuid &      182 &              2.76 \\
          analytics.google.com &                          sid &      490 &              2.71 \\
securepubads.g.doubleclick.net &                          dlt &       66 &              2.57 \\
      www.google-analytics.com &                           dl &      214 &              2.40 \\
   googleads.g.doubleclick.net &                          lmt &      620 &              2.33 \\
                   i.liadm.com &                  bidder\_uuid &       85 &              2.30 \\
\bottomrule
\end{tabular}

}
    \caption{Prevalence of Top-50 \ats link decorations and the number of cookies they share}
    \label{tab:top_50_combined_link_decorations_cookies}
\vspace{-.2in}
\end{table}

\vspace{-5pt}
\subsection{Sharing of Browser Storage through \ats Link Decorations}
\vspace{-5pt}
Prior work has shown that browser storage, such as cookies and local storage, is widely used for tracking by \ats \cite{Siby22WebGraph, Munir23CookieGraphArxiv, Sanchez2021JourneyToCookies}.
To investigate whether \ats link decorations are used to share browser storage, we look for the presence of browser storage in link decorations. 
Table \ref{tab:top_50_combined_link_decorations_cookies} shows the number of distinct browser storage keys shared in the plaintext or encoded formats (Base64, SHA1, SHA-256, and MD5) by top-50 \ats link decorations.
We find that most of the \ats link decorations detected by \name are used to share a large number of browser storage keys (cookies and local storage).  
For example, \texttt{google\_hm} on \textit{cm.g.doubleclick.net} is used by Google's cookie matching service \cite{GoogleRTBCookieGuide} to receive cookies of its advertising partners such as \texttt{CMID} (set by \textit{casalemedia.com}), \texttt{suid} (set by \textit{simpli.fi}), \texttt{\_\_mguid\_} (set by \textit{mediago.io}), and \texttt{tuuid} (set by \textit{adsrvr.com}).
Other such well-known \ats link decorations that are similarly used to share browser storage keys set by various organizations include \texttt{partner\_uid} (LiveRamp), \texttt{partner\_device\_uid} (Tapad), \texttt{uid} (Yahoo!), \texttt{put} (RubiconProject), and \texttt{ttd\_puid} (The Trade Desk). 
We also note that some \ats link decorations are used to share only a few browser storage keys. 
For example, \texttt{fbp} (\textit{www.facebook.com}) is mainly used to share ghosted \cite{Sanchez2021JourneyToCookies,Munir23CookieGraphArxiv} \texttt{\_fbp} cookie set by Meta pixel \cite{fbp-fbc-parameters}.  
As another example, \texttt{vid} and \texttt{sid} (\textit{bat.bing.com}) are mainly used to share ghosted \texttt{\_uetvid} and \texttt{\_uetsid} cookies set by Microsoft UET tracking tag \cite{MicrosoftConversionTracking}.
A related interesting example is \texttt{cid} (\textit{www.google-analytics.com}), which is mainly used to share ghosted \texttt{\_ga} cookie set by Google Analytics \cite{googleciddoc, googlegaparameterreference}. 
However, it is also used to share 250 other browser storage keys potentially due to cookie name conflicts \cite{Zhang20FSEIdentifierConflicts}.

\subsection{Sharing of Deterministic Information through \ats Link Decorations}
Next, we look at the sharing of deterministic identifiers through \ats link decoration.
Deterministic identifiers include email addresses and any commonly used user identifiers such as username, phone number, etc. provided by the user to log into a site.
In contrast with identifiers stored in third-party cookies, these deterministic identifiers are not automatically sent with requests, necessitating their sharing through mechanisms such as link decorations.
To study this sharing of deterministic identifiers, we crawl the 20K sample of top-million websites twice, both with and without providing deterministic identifiers such as email addresses in text input fields.\footnote{To automatically fill in the text fields on web pages, we extended a crawler released by prior work~\cite{senol2022leaky_forms}.}
Similar to the aforementioned storage analysis, we analyze whether link decorations contain these deterministic identifiers in plaintext or encoded format using well-known hashing techniques (Base64, SHA1, SHA256, and MD5).
We repeat these parallel crawls two additional times for a total of three crawls with and without entering deterministic identifiers.
We find 538 link decorations that are present in multiple crawls where we enter deterministic identifiers, but are absent in crawls where we do not enter deterministic identifiers.
\name labels 62 of these 538 link decorations as \ats.

\begin{table}[!t]
    \centering
     	\resizebox{0.8\linewidth}{!}{
      \begin{tabular}{c|l r}
    \toprule
          \textbf{FQDN} &                    \textbf{Key} &  \textbf{Sites} \\
    \midrule
            p.adsymptotic.com & \_expected\_cookie &        90 \\
        idsync.reson8.com &           userid &        47 \\
          api-2-0.spot.im &           ayl\_id &        33 \\
    comcluster.cxense.com &              glb &        29 \\
           polo.feathr.co &           ttd\_id &        21 \\
         rtb.mfadsrvr.com &                \_ &        20 \\
            cs.iqzone.com &             puid &        18 \\
       pixel-geo.prfct.co &              xid &        13 \\
    sync.richaudience.com &         pmUserId &        11 \\
             rp.liadm.com &      ext\_\_pubcid &         8 \\
      cdn-p.cityspark.com &                b &         6 \\
        ssl.connextra.com &          path||0 &         6 \\
             aps.zqtk.net &              url &         5 \\
trackingapi.trendemon.com &         CookieId &         5 \\
     cs-tam.yellowblue.io &              aid &         5 \\
           b6.im-apps.net &              vid &         4 \\
            m.trafmag.com &               id &         4 \\
         sync.upravel.com &              uid &         3 \\
           cdn.gladly.com &                q &         3 \\
           cms.getblue.io &       appnexusid &         3 \\
    \bottomrule
    \end{tabular}
    }
     \caption{Top-20 link decorations which were only present in at least 2 of the crawls where user email address was entered in text fields}
     \vspace{-10pt}
    \label{tab:linkdecoration-unique-to-form-entry-top-10}
\end{table}
Table \ref{tab:linkdecoration-unique-to-form-entry-top-10} shows the top-20 such link decorations.
Most notable of these \ats link decorations are \texttt{ttd\_id} used by Feathr (provides marketing campaign solutions to non-profits \cite{feathr-advertising-solution}), \texttt{pmUserId} used by Rich Audience (an advertising solution which works with major universal identifiers such as by ID5 \cite{id5} and The Trade Desk unified ID \cite{tradedeskunifiedid, rich-audience-solution}), and \texttt{ext\_\_pubcid} used by LiveIntent (provides ``cookieless email-based solutions'' \cite{liveintent-advertiser-solutions}). 

\subsection{Sharing of Probabilistic Information through \ats Link Decorations}
Finally, we look at the exfiltration of probabilistic information through \ats link decorations.
Probabilistic information includes features such as screen resolution, fonts, etc. that can be combined to extract a browser fingerprint \cite{Munir23CookieGraphArxiv, Iqbal21FingerprintingSP}.
To determine if link decoration is potentially being used to send out probabilistic information, we use FP-Inspector an ML-based tool proposed by prior work~\cite{Iqbal21FingerprintingSP} to detect whether the initiator scripts of link decorations are fingerprinters. 
We run FP-Inspector to detect 1,528 fingerprinting scripts on the tested websites.
These fingerprinting scripts initiate requests containing 1,800 unique link decorations, out of which 200 are labeled as \ats by \name.
Table \ref{tab:fingerprinting-decorations} shows the top 20 most common \ats link decorations sent by fingerprinting scripts.
While it is challenging to reverse-engineer the scripts, names (e.g., \texttt{aduid}, \texttt{uuid}) of some of these link decorations make it fairly obvious that they contain some sort of identifiers.
\begin{table}[!t]
    \centering
    \resizebox{\linewidth}{!}{
    \begin{tabular}{c|l r}
    \toprule
    \textbf{FQDN} &                     \textbf{Key} &  \textbf{Sites} \\
    \midrule
securepubads.g.doubleclick.net &                      ippd &              212 \\
             kraken.rambler.ru &                       tid &              146 \\
              sofire.baidu.com &                         t &              127 \\
                  ib.adnxs.com &                   path | 0 &              117 \\
             kraken.rambler.ru &                 top100\_id &              115 \\
             kraken.rambler.ru &                        lv &              104 \\
                 log.rutube.ru &                       sid &               92 \\
             kraken.rambler.ru &                     aduid &               55 \\
     trackingapi.trendemon.com &                       vid &               41 \\
          connect.facebook.net &                   path | 0 &               41 \\
             kraken.rambler.ru &                adtech\_uid &               39 \\
         hexagon-analytics.com &                        uu &               37 \\
                api.segment.io &                   path | 1 &               35 \\
     trackingapi.trendemon.com & MarketingAutomationCookie &               27 \\
             idr.cdnwidget.com &                     bxvid &               24 \\
            events.bouncex.net &                   visitid &               23 \\
              sofire.baidu.com &                         h &               22 \\
              unseenreport.com &                      uuid &               20 \\
             ids.cdnwidget.com &                   path | 0 &               18 \\
         hexagon-analytics.com &                         h &               17 \\
\bottomrule
    \end{tabular}
    }
    \caption{Top-20 \ats link decorations used by scripts involved in fingerprinting and the number of sites they were detected on.}
    \vspace{-10pt}
    \label{tab:fingerprinting-decorations}
\end{table}
\vspace{-10pt}

\section{Discussion}
\label{sec:limitations}
In this section, we further discuss the robustness of \name to evasion, \name's deployment in browsers and browser extensions, and coverage of \name's dynamic analysis. 
\vspace{-10pt}

\subsection{Robustness to Evasion} 
We evaluate the robustness of \name against three different evasion techniques: manipulation of link decoration key names, splitting link decoration values, and combining of link decoration values.

First, we evaluate whether changing link decoration keys impacts \name's accuracy.
To this end, we randomly change the query parameter names as well as the position of resource paths and fragments. 
Our evaluation shows that there is no change in \name's accuracy due to this randomization. 
This is because \name does not directly use name features for query parameters and fragments, and thus there is no impact on the features. 
Moreover, changing the key names of resource paths by changing their position in the URL impacts just one feature (maximum depth of decoration) but does not end up changing the classification outcome for any link decoration.

Second, we evaluate whether splitting a link decoration into multiple link decorations impacts \name's accuracy.
If the individual character lengths of the new smaller link decorations are shorter than 8 characters, \name's pre-processing would remove such link decorations from the classification pipeline -- resulting in a successful evasion.
To mitigate this issue, we can exclude this pre-processing step, but it might result in more false positives.
We evaluate \name (without this pre-processing step) against this evasion technique by splitting the link decorations longer than 8 characters into multiple smaller link decorations. 
Our evaluation shows that splitting link decoration results in only a 0.4\% drop in accuracy, which corresponds to an increase of false positive rate by 0.17\%.

Finally, we evaluate whether combining all link decorations into a single encrypted string impacts \name's accuracy.
This approach has been attempted by Facebook \cite{facebookurlobfuscation} to circumvent query parameter stripping.
In Facebook's case, the obfuscated URL (e.g., {\textit{https://www.facebook.com/user/posts/pfbid0RjTS7KpBA...}}) contains a single encrypted resource path that essentially combines multiple query parameters.
A consequence of combining link decorations in an encrypted string is that \name would be unable to attribute storage exfiltration features to this new link decoration.
However, this change also results in higher entropy and, as discussed in Section \ref{subsec:feature_analysis}, increases the likelihood that the link decoration will be labeled as \ats.
We evaluate \name against this evasion technique by combining the link decorations in an SHA-256 encoded string for 156,348 requests containing both \ats and \nonats link decorations.
Here, \name only uses the features whose values are the same across all the combined link decorations (e.g., number of ancestors, descendants, presence of ad keyword in the ascendant). 
Our evaluation shows that \name is still able to detect 83.4\% of new link decorations as \ats.

\vspace{-5pt}
\subsection{\name's implementation in privacy-focused browsers or browser extensions}
\vspace{-5pt}
While \name's implementation is not suitable for runtime deployment (mainly due to the performance overheads of the browser instrumentation and subsequent dynamic analysis), it can be used in existing privacy-focused browsers and browser extensions as follows.
Concretely, we run \name on live webpages to detect \ats link decorations offline, and then add the detected link decorations to a filter list used in Brave, Firefox, Safari, uBlock Origin, or AdGuard for runtime URL sanitization ~\cite{bravelistqueryparameters, firefoxlistqueryparameters, privacytests, ublockoriginlistqueryparameters, adguardlistqueryparameters}.
We generated a filter list compatible with popular privacy-focused extensions such as uBlock Origin and AdBlock Plus by running \name on our dataset \cite{purl_sanitizer_2021}.
Our filter list is incorporated in the \texttt{adfilt} filter list \cite{adfilt-github}, which is used by AdGuard. 
Note that \name's classifier can be periodically rerun to generate a new filter list in case a tracker frequently changes their link decoration keys/names.
If a tracker randomly generates link decoration keys/names each time, \name's filter list can applied based on their relative position in the URL. 
Note that it is challenging in practice for a tracker to change its link decorations at a fast pace because it requires changing both client-side and server-side logic across different servers and organizations.

\vspace{-5pt}
\subsection{Coverage}
\name builds a graph representation of the webpage execution.
The number of interactions captured depends on the intensity and variety of user activity on a webpage (\eg scrolling activity, number of internal pages clicked). 
\name may not detect certain \ats link decorations if its graph representation does not capture certain interactions between different elements in the webpage because it does not sufficiently emulate different user interactions. 
We attempt to mitigate this issue by randomly scrolling and clicking, but it might not be always sufficient. 
The coverage of \name's dynamic analysis can be improved, if needed, using various techniques from prior research \cite{muzeel,hu2018jsforce}.

\vspace{-5pt}
\section{Conclusion}
\label{sec:conclusion}
In this paper, we investigated the abuse of link decoration for tracking. 
We found that link decoration is used by known trackers for both functional and tracking purposes, even within a single URL, necessitating a fine-grained approach to detect tracking link decorations. 
We proposed \name --- a machine learning approach to detect and sanitize tracking link decorations. 
\name leverages a graph representation that captures interactions and the flow of information across multiple layers of the web stack.

Our evaluation showed that \name significantly outperformed existing countermeasures in its ability to detect link decorations accurately and in minimizing website breakage. 
Our deployment of \name on top-million sites showed that link decoration is abused for tracking on almost three-quarters of the websites by well-known advertising and tracking services to exfiltrate first-party cookies, email addresses, and fingerprints. 
We also showed that \name is robust to common evasion attempts and is readily deployable in privacy-focused browsers and browser extensions as a filter list. 
While \name is orthogonal to existing countermeasures that focus on detecting specific types of tracking, it can be deployed alongside them for a defense-in-depth strategy against new and emerging online tracking techniques.

For reproducibility and to foster follow-up research, \name's source code (OpenWPM patch and the machine learning pipeline) and the detected list of link decorations are available at \cite{purl_sanitizer_2021}.

\bibliographystyle{plain}
\bibliography{bibliography}

\newpage

\appendix
\section{Appendix}
\subsection{Feature Analysis}
\label{subsec:feature_analysis}
We conduct feature analysis to understand which properties of link decorations are most useful in classifying into \ats or \nonats.

First, we look at which feature was the most important when classifying both \ats and \nonats link decorations.
We rank the most important features by summing the feature contributions during the classification of link decorations \cite{treeinterpreter}. 
Table \ref{tab:feature-importance-positive}  reports the top-10 features ranked based on the percentage of instances where the feature was important for \ats classification.
We observe that for \ats instances, Shannon Entropy, the number of nodes, edges, predecessors, and ancestors are the most important.
Figure \ref{fig:shanon-entropy-cdf} plots the conditional distribution of two top-ranked features: Shannon entropy and the number of edges.
Figure \ref{fig:shanon-entropy-cdf} shows that the Shannon entropy for \ats decorations is higher than \nonats decorations.
Specifically, more than 70\% of \nonats link decorations and only 8\% of \ats link decorations have Shannon entropy lower than 3, respectively.
The usage of higher entropy strings by \ats link decorations is expected as they are more suitable for storing unique identifiers.
Figure \ref{fig:num-edges-cdf} shows that \ats decorations are more connected as compared to \nonats decorations.
Specifically, less than 50\% of \ats link decorations and more than 85\% of \nonats link decorations have less than 10,000 edges, respectively. 
\ats link decorations tend to interact more with other elements of the webpage, which is expected as \ats link decorations are expected to be fetched and updated more frequently from storage and shared more frequently through network requests.

\begin{figure}[!htpb]
    \centering
    \includegraphics[width=0.8\linewidth]{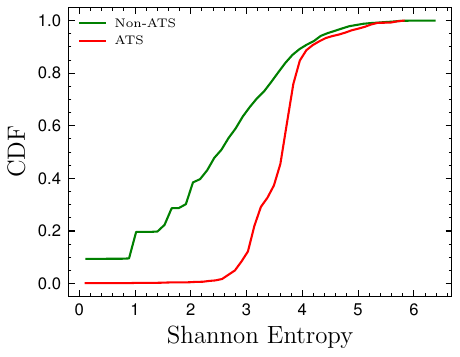}
    \caption{Distribution of Shannon entropy value for \ats and \nonats link decorations. \ats link decorations have a higher Shannon entropy as compared to \nonats link decorations.}
    \label{fig:shanon-entropy-cdf}
\end{figure}
\begin{figure}[!htpb]
    \centering
    \includegraphics[width=0.8\linewidth]{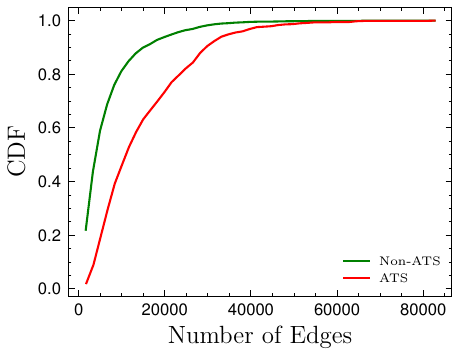}
    \caption{Distribution of the number of edges for \ats and \nonats link decorations. \ats link decorations interact more with the neighboring nodes, resulting in higher connectivity and number of edges than \nonats link decorations.}
    \label{fig:num-edges-cdf}
\end{figure}

In addition to these features, the flow of information from storage nodes to link decoration nodes also affects the decision due to the importance of indirect features, which are calculated by exfiltration and infiltration to and from the storage nodes.
We find that \ats link decorations are more likely to exfiltrate storage values, with an \ats decoration averaging 7.4 exfiltrations (standard deviation of 20.56) while a \nonats decoration averaging only 0.06 (standard deviation of 1.41).
\begin{figure}[!htpb]
    \centering
    \includegraphics[width=0.8\linewidth]{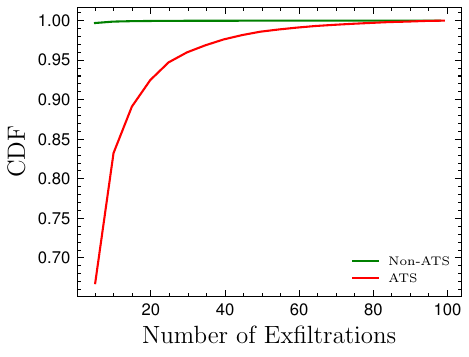}
    \caption{Distribution of the number of exfiltrations for \ats and \nonats link decorations. \ats link decorations are used to exfiltrate storage values significantly more than \nonats link decorations.}
    \label{fig:num-exfil-cdf}
\end{figure}

For \nonats instances, we observe that lack of information flow was a strong indicator, with the top three features directly or indirectly related to exfiltration and infiltration of storage values.
The classifier also took into account the entropy of the string and keywords related to advertisement in ascendants as important features for classifying \nonats link decorations.
Our analysis shows that flow features, that map the flow of information from storage to link decorations, entropy, and the position of link decoration within the URL are the most important features when it comes to classifying link decorations into \ats and \nonats categories.

\begin{table}[!ht]
    \centering
    \begin{tabular}{l r}
    \toprule
    \textbf{Feature}   & \textbf{Percentage}  \\
    \midrule
    Shannon Entropy         &             17.43\% \\
    Number of Nodes                &            16.38\% \\
    Closeness Centrality of Indirect Edges &   13.22\% \\
    Ratio of Nodes over Edges        &            12.4\% \\
    Number of Edges                  &          11.54\% \\
    Number of Script Predecessors    &           5.28\% \\
    Number of Ancestors              &               3.22\% \\
    Ratio of Edges over Nodes      &              3.06\% \\
    Number of Indirect Ancestors   &             2.83\% \\
    Closeness Centrality      &            2.03\% \\
    \bottomrule
    \end{tabular}
    \caption{Percentage of instances where a feature was the most important for \ats link decoration classification}
    \label{tab:feature-importance-positive}
\end{table}

\subsection{\name Features}
\label{app:features}
\begin{table}[!ht]
	\centering
	\resizebox*{\columnwidth}{!}{%
		\begin{tabular}[c]{l c}
			\toprule
			\textbf{Feature}                                                        & \textbf{Type} \\
			\midrule
			Graph size (\# of nodes, \# of edges, and nodes/edge ratio)             & Structure     \\
			Degree (in, out, in+out, and average degree connectivity)               & Structure     \\
			Centrality (closeness centrality, eccentricity)                         & Structure     \\
			Ascendant’s attributes                                                  & Structure     \\
			Descendant of a script                                                  & Structure     \\
			Ascendant’s script properties (Ad keyword, FP keyword, length of script)                                           & Structure     \\
			Parent is an eval script                                                & Structure     \\
            \midrule
			Depth of Link Decoration in URL                                                & Content     \\
			Shannon Entropy                                                & Content     \\
			\midrule
			Local storage access by parent (\# of sets, \# of gets)                           & Flow          \\
			Cookie accesses by parent (\# of sets, \# of gets)                                  & Flow          \\
			Requests (sent, received) by parent                                               & Flow          \\
			Redirects (sent, received, depth in chain) by parent                             & Flow          \\
			Common access to the same storage node                                  & Flow          \\
			Cookie exfiltration                                                     & Flow          \\
			Cookie infiltrations by parent                                                     & Flow          \\
			Cookie Setter (\# of exfiltration, \# redirects) by parent                        & Flow          \\
			Graph size (\# of nodes, \# of edges, and nodes/edge ratio)             & Flow          \\
			Degree (in, out, in+out, and average degree connectivity)               & Flow          \\
			Centrality (closeness centrality, eccentricity)                         & Flow          \\
			\bottomrule
		\end{tabular}
	}
	\caption{Features used by \name. \name calculates Graph size, Degree, and Centrality features using both normal and shared information edges. The former comes under structural features while the latter comes under flow features.}
	\label{tab:feature_comparison}
\end{table}

\end{document}